\documentclass[showpacs,preprintnumbers,amsmath,amssymb,11pt]{revtex4}
\pdfoutput=1                                     %use pdflatex file.tex
\usepackage{dcolumn}                             % Align table columns on decimal point
\usepackage{dsfont}
\usepackage{slashed}
\usepackage{graphicx}                            % Include figure files
\usepackage{epsfig}
\usepackage{longtable}                           % For table that spans more than one page or column
\usepackage{color}

\usepackage{subfigure}
\usepackage{bm}                                  % Bold math
\usepackage{bbm}
\usepackage{hyperref}                            % References as links

\def\cyp{a}
\def\cyi{b}
\def\juelich{c}
\def\mit{d}
\def\hna{e}
\def\ntua{f}
\newcommand{\imag}{\mbox{i}}   
\newcommand{\be}{\begin{equation}}
\newcommand{\ee}{\end{equation}}
\newcommand{\beq}{\begin{eqnarray}}
\newcommand{\eeq}{\end{eqnarray}}
\newcommand{\nn}{\nonumber}
\begin{document}
\title{Nucleon to $\Delta$ transition form factors with $N_F=2+1$ domain wall 
fermions}
\author{C. Alexandrou$^{(\cyp,\cyi)}$, G. Koutsou$^{(\juelich)}$,  J. W. Negele$^{(\mit)}$,  Y. Proestos$^{(\cyi)}$ and A. Tsapalis$^{(\hna, \ntua)}$}
%\email{alexand@ucy.ac.cy, i.koutsou@fz-juelich.de,negele@mit.edu, proestos@cyi.ac.cy, a.tsapalis@iasa.gr }
\affiliation{
 {$^{(\cyp)}$ Department of Physics, University of Cyprus, P.O. Box 20537, 1678 Nicosia, Cyprus}\\
{$^{(\cyi)}$  Computation-based Science and Technology Research Center, The Cyprus Institute, P.O. Box 27456, 1645 Nicosia, Cyprus }\\
 {$^{(\juelich)}$ Department of Physics, University of Wuppertal/Forschungszentrum J\"{u}lich D-52425, J\"{u}lich, Germany}\\
{  $^{(\mit)}$ Center for Theoretical Physics, 
Laboratory for
Nuclear Science and Department of Physics, Massachusetts Institute of
Technology, Cambridge, Massachusetts 02139, U.S.A.}\\
{$^{(\hna)}$  Hellenic Naval Academy, Hatzikyriakou Ave., Pireaus 18539, Greece} \\
{$^{(\ntua)}$ Department of Physics, National Technical University of 
Athens, Zografou Campus 15780, Athens, Greece}
}
\begin{abstract}

We calculate the electromagnetic, axial and 
pseudo-scalar form factors of the Nucleon to $\Delta(1232)$ transition
using two dynamical light degenerate quarks and a dynamical strange quark 
simulated with the domain wall fermion action. Results
are obtained at  lattice spacings
$a = 0.114$~fm  and $a=0.084$~fm,  with corresponding pion masses of 
$330$~MeV and $297$~MeV, respectively.
High statistics measurements are achieved by utilizing the coherent sink 
technique. The dominant electromagnetic dipole form factor, the
 axial 
form factors and the pseudo-scalar coupling
  are extracted to a good accuracy. This allows  the investigation 
 of the  non-diagonal 
Goldberger-Treiman relation.
 Particular emphasis is given on the extraction of the sub-dominant 
electromagnetic quadrupole form factors  and their ratio to
the dominant dipole form factor, $R_{EM}$ and $R_{SM}$, measured in experiment.
\end{abstract}
\maketitle
\section{\label{sec:intro} Introduction}

Form factors are fundamental quantities which probe
the  internal structure of the hadron. 
They are typically extracted from electromagnetic or weak 
scattering processes on hadronic targets,  dominated 
by one-body exchange currents.
The prime example are the
form factors of the proton,  which 
 remain the  most well-studied.
Its electromagnetic (Sachs) form factors have been measured since the 50's~\cite{Gao:2003ag} and
static properties such as the magnetic moment and the charge radius
are extracted. For recent reviews on the
experimental and theoretical status  we refer the reader to Refs.~\cite{Gao:2003ag,Hyde-Wright} and \cite{Perdrisat:2006hj,Arrington:2006zm} respectively. Despite
the long history of measurements of the electromagnetic nucleon form factors, polarization experiments
recently revealed an unexpected behaviour in  the momentum dependence 
of the electric to magnetic form factor of the proton which has 
triggered theoretical investigations to explain the  dynamics that give rise to such behavior~\cite{Afanasev:2005mp}.

The proton, being the building block of all matter that is presently observed to be stable,
provides a nice laboratory for studying a relativistic bound state.
One fundamental question is whether hadrons being composite systems are deformed
and in particular whether the proton is spherical or has an intrinsic deformation.  The elastic form factors do not suffice to answer this question on
 nucleon deformation, an important quantity that characterizes the distribution
of quarks in the nucleon.
The reason lies in the fact that the spectroscopic quadrupole moment of an $J = 1/2$ state vanishes
identically in the laboratory frame if a one-photon exchange process is studied, although a
quadrupole deformation may still exist in the body-fixed intrinsic frame. Therefore, regarding the
nucleon, one has to study the transition to the lowest positive parity $J=3/2$ state which is the
$\Delta(1232)$. The $\gamma N \Delta$ matrix element is parameterized in terms of a dominant
magnetic dipole, $G_{M1}$, plus the sub-dominant electric quadrupole, $G_{E2}$, and Coulomb quadrupole,
$G_{C2}$, transition form factors. Detection of non-zero $G_{E2}$ or $G_{C2}$ signals the existence
of deformation in the $N-\Delta$ system~\cite{Papanicolas:2007zz,Bernstein:2007jt,Papanicolas:2003zz}. Precise  electroproduction experiments in the last
decade demonstrated that this is indeed the case and provided measurements of the EM transition form
factors for a wide range of values of the  momentum transfer squared $q^2$.
The $E2$ and $C2$ amplitudes are measured
to a few percent of the dominant, $M1$, amplitude and  are typically given
 as  ratios  to the $M1$ amplitude, denoted by $R_{EM}$ and $R_{SM}$ respectively.

State-of-the-art lattice QCD calculations can yield model independent 
results on hadron form factors,
thereby  providing direct comparison with experiment.
Like in experiment, the electromagnetic nucleon form factors
have been studied by many collaborations
recently using dynamical 
simulations~\cite{Alexandrou:2010,Yamazaki:2009zq,Syritsyn:2009mx,Bratt:2010jn,Alexandrou:2009ng,Alexandrou:2006ru,Ohta:2008kd}. 
Reproducing the  experimental results on the electric and magnetic form factors is a prerequisite for enabling lattice predictions of other form
factors. This is also true for lattice calculations of the dominant 
magnetic dipole N to $\Delta$ transition form factor which is also well measured experimentally. In particular, in the case of the N to $\Delta$, there are no
disconnected contributions and therefore reproducing this form factor
would provide a validation of lattice QCD techniques
in calculating hadron form factors. 
The evaluation of the
 sub-dominant N to $\Delta$ electric and Coulomb quadrupole form factors
have also been studied for many years in dedicated experiments since, as we
already pointed out, a non-zero
value of these form factors signals a deformation in the N-$\Delta$ system.
However the experimental determination needs model input and therefore
lattice QCD can provide an ab initio calculation of these fundamental quantities. 

In the axial sector, in the case of the nucleon, there exist two form factors, the
axial, $G_A$, and induced pseudo-scalar, $G_p$, form factors. They
 have been studied in neutrino scattering
and muon capture experiments, respectively but experimental data are 
less precise~\cite{Gorringe:2002xx,Bernard:2001rs}. 
 There have also been  
 several lattice evaluations 
of the nucleon axial charge $g_A$~\cite{Alexandrou:2010,Edwards:2005ym, Khan:2006de, Yamazaki:2008py} and of the momentum dependence of the two form factors~\cite{Alexandrou:2007xj,Bratt:2010jn}. Partial conservation of axial symmetry (PCAC)
leads to a relation between the nucleon axial charge and the pseudo-scalar
$\pi-N$ coupling constant $g_{\pi NN}$,
  the well-known Goldberger-Treiman
relation.
The strong decay of the $\Delta$ 
obscures greatly experimental studies of the N to $\Delta$ 
weak matrix element but some information on the dominant axial transition form factors 
$C_5^A(q^2)$ and $C_6^A(q^2)$ is available from  neutrino interactions on hydrogen and
deuterium targets. $C_5^A$ and $C_6^A$ are the analogue of 
 the nucleon axial form factors, $G_A$ and
$G_p$, respectively. Indeed, like $G_p$, the $q^2$
dependence of $C_6^A$  is dominated by the pion pole 
and due to the axial 
Ward-Takahashi identity (AWI) a relation can 
be derived between $C_5^A$ and the phenomenological strong coupling of the pion-nucleon-$\Delta$ vertex,
$g_{\pi N \Delta}$. This relation is referred to as the non-diagonal Goldberger-Treiman relation.

Such observations strongly motivate the study of the $N$-to-$\Delta$ transition from first principles
using lattice QCD.
 The first lattice study of the electromagnetic $\gamma N \Delta$ 
transition was carried out in the quenched approximation~\cite{Leinweber:1993} 
 at a fixed Euclidean momentum transfer squared $Q^2=-q^2$ with inconclusive results as to whether the $E2$ or $C2$ amplitudes were non-zero due to large statistical errors.
A study employing the formalism of Ref.~\cite{Leinweber:1993} followed
using quenched and two dynamical
flavors of degenerate Wilson-type quarks at smaller quark masses but still only at the lowest $q^2$-value allowed on the lattices at hand.
Although there was an almost ten-fold increase in statistics
the  values obtained for the quadrupole form factors had large statistical 
noise and a zero value could not be excluded~\cite{Alexandrou:2003ea,Alexandrou:2003ab}.  In order to obtain  sufficient
 accuracy we combined sequential 
inversions through the source instead of through the current 
for the evaluation of the three-point functions and 
 optimized sources that led to a large sample of 
statistically independent measurements for a given $q^2$-value. The calculation,
carried out in the quenched approximation, confirmed a non-zero value with the 
correct sign for both of the quadrupole amplitudes~\cite{Alexandrou:2004xn,Alexandrou:2004fr}. 
A similar study was also carried out for the axial vector N to $\Delta$ matrix element~\cite{Alexandrou:2006mc}. 
Using this new methodology we extended the calculation of the $N$ to $\Delta$ 
electro-weak form factors to unquenched lattice QCD. For the latter study we
used $N_f=2$ Wilson fermions as well as an $N_f = 2+1$ calculation with a mixed action  with domain
wall valence quarks on a staggered sea reaching a pion mass of about 350 
MeV~\cite{Alexandrou:2007dt,Alexandrou:2007xj,Alexandrou:2005em,Alexandrou:2007hr}. This  calculation showed that the unquenched results on the Coulomb 
quadrupole form factor at low $q^2$ decreased towards the
experimental results. However, the discrepancy in the momentum dependence
of the dominant dipole form factor remained with lattice results having
smaller values at low $q^2$-values and
a weaker dependence on $q^2$. Using the same set of sequential propogators
as in the electromagnetic case
the axial and pseudo-scalar  $N$ to $\Delta$ form factors were
 studied~\cite{Alexandrou:2007xj,Alexandrou:2007hr}. 
The strong coupling constant 
 $g_{\pi N \Delta}$ and non-diagonal Goldberger-Treiman relation were examined 
in detail and it was demonstrated that the behaviour is  very similar manner to the corresponding relations in the nucleon system.

In this work we study   the $N$-to-$\Delta$ transition using $N_F=2+1$ dynamical
domain wall fermions  simulated by the RBC-UKQCD collaborations~\cite{Allton:2008pn}.
This eliminates ambiguities about the correctness of the continuum limit due to the rooting of the staggered sea quarks and the matching
required in a mixed action.
Preliminary results have been presented in Refs.~\cite{Alexandrou:2009uc}. 
We use two ensembles corresponding to lattice spacing  $a=0.114$~fm and $a=0.084$~fm
and  physical volume
of $(2.7 \;{\rm fm})^3$.  Both lattice spacings are smaller than the lattice spacing used in our previous mixed-action calculation.
This allows, for the first time, 
 the investigation of  cut-off effects on these hadronic
observables.
For each lattice spacing, we chose to perform the calculation on the lightest
pion mass set available, namely at  330 MeV pions for the coarse lattice and 297 MeV for the fine one, in order to be as close as possible to  the physical regime. The goal is, first, to check whether lattice results on
 the well measured experimentally dominant dipole form approach experiment. 
Secondly, we would like
to see the onset of the large pion cloud contributions to the quadrupole form factors as predicted by chiral effective theory~\cite{Pascalutsa:2005}. 
Thirdly we will extract the axial N to $\Delta$  coupling that enters in chiral
expansions of the nucleon axial charge as well as
the strong coupling constant $g_{\pi N\Delta}$. Determining these
quantities together with the corresponding quantities
$g_A$ and $g_{\pi NN}$ for the nucleon as well as for the $\Delta$ on the same gauge configurations will enable simultaneous chiral extrapolations to the physical point and yield more reliable results on these fundamental quantities.

The paper is organized as follows: In Section~\ref{sec:latt-tech} we describe the general lattice setup 
and outline the techniques utilized to extract all the transition form 
factors from three-point functions measured on the lattice. In 
Section~\ref{sec:EM-ffs} we present in detail the decomposition of the electromagnetic 
N to $\Delta$ matrix element on the hadronic level in terms of the Sachs form factors and 
discuss the results for the electromagnetic transition form factors. 
In Section~\ref{sec:AX-ffs} we give the corresponding matrix element for the electro-weak transition 
and discuss the results on the axial and pseudo-scalar form factors. 
Finally, the last section contains our conclusions and  an outlook regarding  further studies in the subject.

\section{\label{sec:latt-tech} Lattice setup and techniques}

We use the $N_f = 2+1$ dynamical domain wall fermion (DWF) ensembles
generated by the  RBC and UKQCD collaborations~\cite{Allton:2008pn,Aoki:2007xm,Scholz:2008uv} with the strange quark mass
fixed at the physical point.
Specifically, we consider gauge configurations on lattices of volume $24^3\times 64$ 
corresponding to a pion mass of about  $330$~MeV and inverse lattice spacing $a^{-1}= 1.73(3)$~GeV and  $32^3\times 64$ corresponding to a pion mass of about  $297$~MeV and $a^{-1}= 2.34(3)$~GeV.
We refer to the former lattice corresponding to $a^{-1}= 1.73(3)$~GeV,  as the \emph{coarse} DWF lattice, and the one corresponding to  $a^{-1}= 2.34(3)$~GeV,  as the \emph{fine} DWF  lattice. 

Domain wall fermions  preserve chiral symmetry in the infinite limit of the 
fifth dimension, $L_5$. In actual computations $L_5$ is finite
 leading to  an additive contribution to the quark mass as defined
through the Axial Ward-Takahashi Identity (AWI). For the coarse ensemble a
residual quark mass of $a m_{res} = 0.00315(2)$ has been measured by 
UKQCD-RBC~\cite{Allton:2008pn} with the extent of the fifth dimension set to
$L_5 = 16$. The same $L_5$ extent for the fine ensemble leads to a much smaller
violation, measured to $a m_{res} = 0.000665(3)$, or just 
$17 \%$ of the bare quark mass~\cite{Syritsyn:2009mx}.

Details about the lattice parameters used in this study are provided in Table~\ref{Table:params_DWF_hybrid}, where for comparison the relevant values of the parameters used
in our previous study using the mixed action~\cite{Alexandrou:2007dt,Alexandrou:2007xj,Bratt:2010jn} are also given.
%
%lattice parameters table
\begin{widetext}
\begin{table}[ht]
\scriptsize
\begin{center}
\begin{tabular}{cccccccccc}
\hline\hline 
Volume & $N^{\mathrm{dom.}}_{\mathrm{confs}}$ ($N_{\mbox{\scriptsize meas.}}$) &  $N^{\mathrm{subd.}}_{\mathrm{confs}}$ ($N_{\mbox{\scriptsize meas.}}$)  &  $a^{-1}$ [GeV] & $Z_V$ & $Z_A$ & $m_{u,d}/m_s$ & $m_\pi$ [GeV] &  $m_N$  [GeV]& $m_\Delta$ [GeV]\\ 
\hline\hline
\multicolumn{7}{c}{coarse $N_F=2+1$ DWF~\cite{Allton:2008pn}}\\
\hline
 $24^3\times 64$ & 200 (800) & 398 (1592)   & 1.73(3)& 0.7161(1) &0.7161(1) &  0.005/0.04 & 0.329(1) &   1.130(6) & 1.457(11)\\
\hline
\multicolumn{7}{c}{fine $N_F=2+1$ DWF~\cite{Syritsyn:2009mx}}\\
\hline
$32^3\times 64$ & 176 (704) & 309 (1236)  & 2.34(3) & 0.7468(39) &0.74521(2)  & 0.004/0.03 & 0.297(5) &  1.127(9) & 1.455(17) \\
\hline
\multicolumn{7}{c}{Hybrid action~\cite{Bratt:2010jn}} \\
\multicolumn{7}{c}{
DWF valence: $ am_{u,d} = 0.0138$, $am_s=0.081$}\\
\hline
$28^3\times 64$ & 300 (300) &300 (300) &   1.58(3) & & &  0.01/0.05 & 0.353(2) &1.191(19) &  1.533(27) \\
\hline\hline
\end{tabular}
\end{center}
\normalsize
\caption{Parameters for the calculation of the electromagnetic and axial transition form factors. The mixed action results from Refs.~\cite{Alexandrou:2007dt,Alexandrou:2007xj} are also included for completeness. 
In the second (third) column we show the  number of gauge configurations used for the  dominant (suppressed) form factors. 
For the DWF lattices the numbers in the parentheses next to the number of configurations  are multiplied by four, since the 
coherent sink method was employed,  showing the actual number of measurements taken into account in the overconstrained analysis. 
In the fifth and sixth columns we list the values of the vector and axial current renormalization constants, respectively that 
have been used as input parameters in our calculation, since we have used local currents and not the lattice conserved ones. }
\label{Table:params_DWF_hybrid}
\end{table}
\end{widetext}

In order to create the proton  and $\Delta^+$ states we use the standard interpolating operators
\be
\chi ^p (x) = \epsilon^{a b c}\; \left[ u^{T\; a}(x)\; C \gamma_5
d^b(x) \right]\; u^c(x) ,
\ee
\be
\chi ^{\Delta^{+}}_\sigma  (x) = \frac{1}{\sqrt{3}} \epsilon^{a b c} \Big
\lbrace
2 \left[ u^{T a}(x)\; C \gamma_\sigma d^b(x) \right]u^c(x) \;
+\; \left[ u^{T a}(x)\; C \gamma_\sigma u^b(x) \right]d^c(x) \Big \rbrace,
\ee
respectively. The $J=3/2$ $\Delta$ state is described by the Rarita-Schwinger vector-spinor 
where $\sigma = 1,2,3,4$ is the Lorentz vector field index. $C=\gamma_4  \gamma_2$ is the charge-conjugation matrix.

Form factors of the $N-\Delta$ transition are extracted on the lattice
from the three-point function
\be 
\langle G_{\sigma}^{\Delta J_\mu N} (t_2, t_1 ;
{\bf
p}^{\;\prime}, {\bf p}; \Gamma_\tau) \rangle  = 
\sum_{{\bf x}_2, \;{\bf
x}_1} e^{-i {\bf p}^{\prime} \cdot {\bf x}_2 } e^{+i {\bf q}
\cdot {\bf x}_1 } \; \Gamma_\tau^{\beta \alpha}
\langle \Omega | 
T\left[\chi_{\Delta}^{\sigma \alpha} ({\bf x}_2,t_2) J_\mu({\bf
x}_1,t_1) \bar{\chi}_N^{\beta} ({\bf 0},0) \right] |\Omega
\rangle  
\label{3pt} 
\ee
In this notation, an initial nucleon state with momentum ${\bf p}$ is created at time zero and propagated to a later time $t_1$ at  which it couples 
to the current $J$  
 causing a transition to the $\Delta$  state  of momentum ${\bf p}^\prime$
which is annihilated at a later time $t_2$.  ${\bf q} = {\bf p}^{\prime}-{\bf p}$ is the momentum transfer. The projection matrices $\Gamma_{\tau}$ are given by

\be
\Gamma_i = \frac{1}{2}
\left(\begin{array}{cc} \sigma_i & 0 \\ 0 & 0 \end{array}
\right), \;\;\;\;
\Gamma_4 = \frac{1}{2}
\left(\begin{array}{cc} \mathbbm{1} & 0 \\ 0 & 0 \end{array}
\right) .
\ee
The one-body currents considered in this work include the local vector current
\be
V_\mu(x) = \frac{2}{3} \bar{u}(x) \gamma_\mu u (x) -\frac{1}{3} \bar{d}(x) \gamma_\mu d(x)\quad,
\label{veccurr}
\ee
the axial-vector current and pseudo-scalar density
\be
A_{\mu}^a(x)= \bar{\psi}(x)\gamma_\mu \gamma_5\frac{\tau^a}{2}\psi(x) \;\;\; , \;\;\;\;
P^a(x)= \bar{\psi}(x)\gamma_5 \frac{\tau^a}{2}\psi(x) 
\label{currents}
\ee
with $\tau^a$ the three Pauli-matrices acting in flavor space
and $\psi$ the  isospin doublet quark field. Note that due to the $\Delta J = 1$ nature of the transition, only the
isovector part of $V_\mu$ contributes and, due to isospin symmetry, only the flavor diagonal operator $\tau^3$ needs to be evaluated.
Inclusion of baryon states in the three-point function~(\ref{3pt}) and the use of standard Euclidean spin-sums for the 
Rarita-Schwinger field
\be
\sum_s u_\sigma(p,s)\bar{u}_\tau(p,s) = \frac{-i\gamma \cdot p+m_\Delta}{2m_\Delta} \left[\delta_{\sigma\;\tau} +\frac{2p_\sigma p_\tau}{3m_\Delta^2}
-i\frac{p_\sigma\gamma_{\tau}-p_\tau\gamma_{\sigma}}{3m_\Delta}
-\frac{1}{3}\gamma_\sigma \gamma_\tau \right] ,
\ee 
and the Dirac spinor  
\be
\sum_s u(p,s)\bar{u}(p,s) =\frac{-i\gamma \cdot p+m_N}{2m_N} 
\ee
lead to the isolation of the desired matrix element, assuming that the initial and final ground states dominate the propagation
before and after the operator  insertion, respectively. In order to cancel,
in the large Euclidean time limit, the dependence on the Euclidean time evolution  and on the unknown
overlaps of the nucleon and $\Delta$ states with the initial states,
 we form the following ratio:
%
%\small 
\begin{align}
%\vspace*{-0.8cm} 
R^{J}_\sigma (t_2, t_1; {\bf p}^{\; \prime}, {\bf p}\; ; \Gamma_\tau ; \mu)
=&  \frac{\langle G^{\Delta J_\mu N}_{\sigma} (t_2,
t_1 ; {\bf p}^{\;\prime}, {\bf p};\Gamma ) \rangle \;} {\langle
G^{\Delta \Delta}_{ii} (t_2, {\bf p}^{\;\prime};\Gamma_4 ) \rangle
\;}  \nn\\
& \times \biggr [\frac{\langle G^{\Delta \Delta}_{ii} (t_2, {\bf
p}^{\;\prime};\Gamma_4 ) \rangle}{ \langle
G^{N N} (t_2, {\bf p};\Gamma_4 ) \rangle }\> 
\frac{ \langle G^{N N}(t_2-t_1, {\bf
p};\Gamma_4 ) \rangle \;\langle G^{\Delta \Delta}_{ii} (t_1, {\bf
p}^{\;\prime};\Gamma_4 ) \rangle} {\langle G^{\Delta \Delta}_{ii}
(t_2-t_1, {\bf p}^{\;\prime};\Gamma_4 ) \rangle \;\langle
G^{N N} (t_1, {\bf p};\Gamma_4 ) \rangle} \biggr ]^{1/2} 
\label{ratio}
\end{align}
\normalsize
which requires also measurements of the  nucleon ($ G^{NN}$) and $\Delta$ ($G^{\Delta \Delta}_{\sigma\tau}$) two-point functions
\begin{align}
\langle G^{NN} (t, {\bf p} ; \Gamma) \rangle &=  \sum_{{\bf x}}
e^{-i {\bf p} \cdot {\bf x} } \; \Gamma^{\beta \alpha}\;\langle \Omega |\;T\;\chi^{\alpha}({\bf x},t) 
 \bar{\chi}^{\beta} ({\bf 0},0)  
\; |  \Omega\;\rangle,  \\
\langle G^{\Delta \Delta}_{\sigma\tau} (t, {\bf p} ; \Gamma) \rangle &=  \sum_{{\bf x}}
e^{-i {\bf p} \cdot {\bf x} } \; \Gamma^{\beta \alpha}\;\langle \Omega |\;T\;\chi^{\alpha}_{\sigma}({\bf x},t) 
 \bar{\chi}^{\beta}_{\tau} ({\bf 0},0)  
\; | \Omega \;\rangle .
\label{NN-DD}
\end{align}
Implicit summations on indices $i=1,2,3$ are assumed in the above ratio~(\ref{ratio}), which is designed such that the time evolution (and consequently 
the noise) 
appearing  in its two-point function part is minimized.
In  the large Euclidean time limit ($t_2 -t_1 \gg 1, \ t_1 \gg 1$) where we have
ground state dominance
this ratio~(\ref{ratio}) thus yields a 
a time-independent function $\Pi^J_{\sigma}({\bf p}^{\; \prime}, {\bf p}\; ;\Gamma_\tau ; \mu) \;$ that is related to the matrix element
$\langle \Delta({\bf p}^\prime)|J|n({\bf p}) \rangle $. Therefore
we look for the plateau region of Eq.~(\ref{ratio}) in order to extract the
matrix element that we are interested in.
For a given operator insertion $J$ and projection matrix $\Gamma_\tau$, the function $\Pi^J_{\sigma}({\bf p}^{\; \prime}, {\bf p}\; ;\Gamma_\tau ; \mu) \;$ 
is a linear combination of the corresponding form factors. These relations
for the appropriate choice of $\Gamma_\tau$ and $\Delta$ vector index $\sigma$ will be given in the following sections.

The computationally intensive part of the calculation lies in the calculation
of the three-point function given in Eq.~(\ref{3pt}). In order to achieve the
extraction of the momentum dependence of the matrix element for the $V_\mu(x)$, $A^3_\mu(x)$ and $P^3(x)$ insertions, one needs an evaluation for
a large number of values of the momentum transfer ${\bf q}$.
This is feasible by evaluating the
matrix element using  {\it sequential inversions through the sink}. 
In this method, the
quantum numbers of the source and sink interpolating fields are fixed, 
effectively by fixing the $\sigma$ and $\tau$ indices. The
time slices of the source and sink are, in addition, fixed. The quark propagator with the operator insertion is obtained by
the joining of a forward propagator
and the {\it sequential propagator} which is obtained by using as a source
 the baryon state at the sink folded in with the two forward propagators 
from the source. 
With the forward and sequential propagators available,
the operator insertion at selected intermediate times $t_1$ and momenta transfers ${\bf q}$ is readily available. In this method the final
state, in this case the $\Delta$-state, is always at rest.
Since the $\sigma-\tau$ space of indices still spans a set of 16 independent  inversions that would be required,
an optimization in this space has been exploited.
Three linear combinations are constructed from which the EM, axial and pseudo-scalar 
form factors are extracted such that the maximal set of statistically independent 
measurements of momentum transfer vectors {\bf q} per $q^2$ value is achieved. 
In addition, they are chosen to decouple the dominant dipole ($M1$) 
part of the EM transition from the sub-dominant quadrupoles $E2$ and $C2$ measurements.
The three linear combinations which we construct and measure in this work are given below. 
\begin{align} 
\label{S123}
S^J_1({\bf q};J)=&\sum_{\sigma=1}^3 \Pi^J_\sigma({\bf 0},-{\bf q}\; ;\Gamma_4 ;J)  \\
S^J_2({\bf q};J)=&\sum_{\sigma\neq k=1}^{3} \Pi^J_\sigma({\bf 0},-{\bf q}\; ;\Gamma_k ;J)  \\  
S^J_3({\bf q};J)=&\;\;\Pi^J_3({\bf 0},-{\bf q}\; ;\Gamma_3 ;J)- \frac{1}{2}\Big[
 \Pi^J_1({\bf 0},-{\bf q}\; ;\Gamma_1 ;J) + \Pi^J_2({\bf 0},-{\bf q}\; ;\Gamma_2 ;J) \Big] \quad,
\end{align} 
where  $J$ denotes the operators $V_\mu$, $A_\mu^3$ and $P^3$.
Occasionally we refer to $S_1,S_2,S_3$ as 
{\it optimal $\Delta$ sinks}, although they actually correspond to an optimal linear combination of the full $N-\Delta$ 
three-point function with arbitrary insertion $J$. We stress that, given the forward propagators, 
{\it three} inversions in total  are required in order to compute the momentum dependence of the
full $N-\Delta$ transition and extract the electromagnetic, axial and pseudo-scalar form factors.

\begin{figure}[htb]
\begin{center}
\hspace{-0.2cm}\includegraphics[width=0.85\linewidth,height=0.75\linewidth]{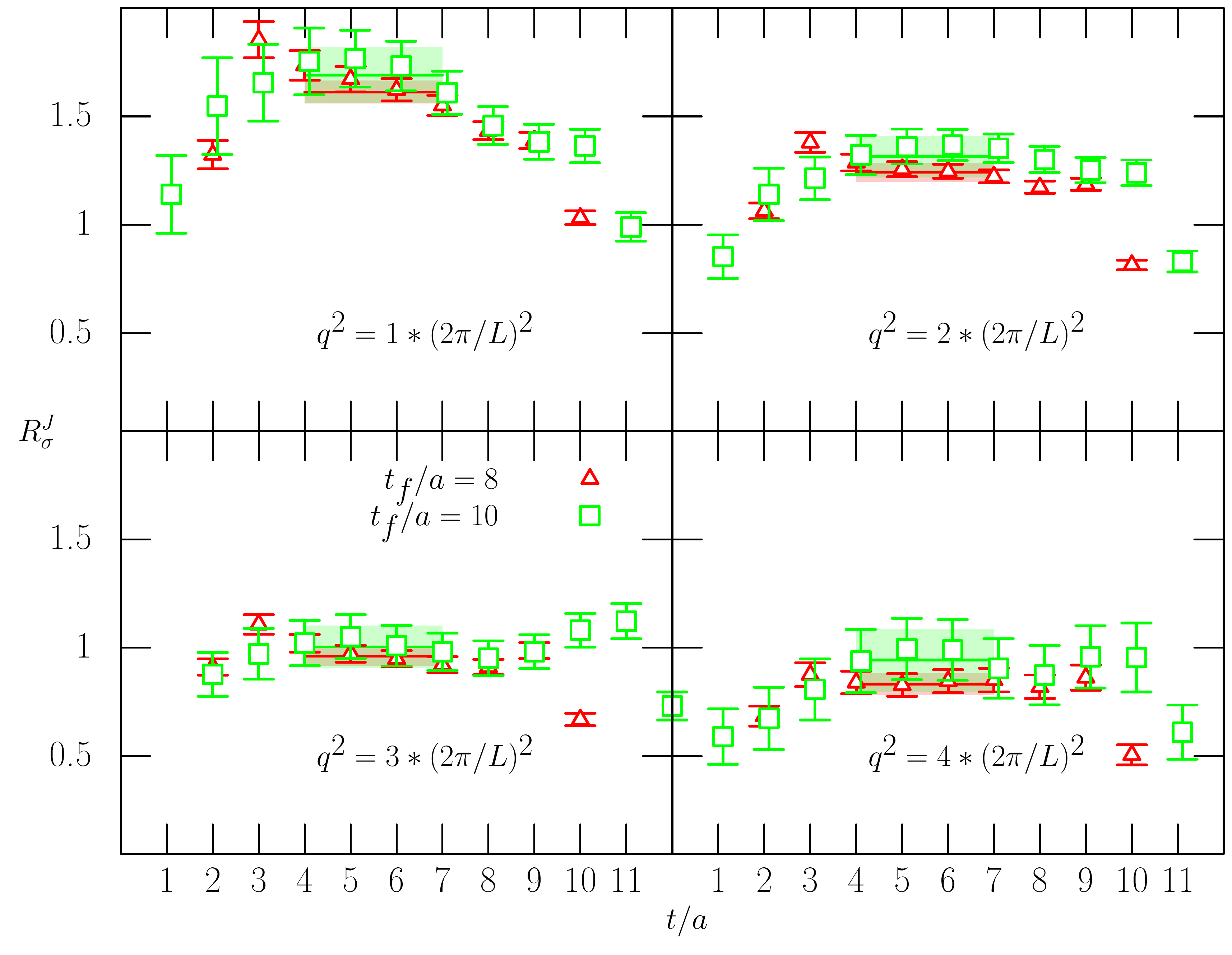} 
\vspace*{-0.5cm}
\caption{
The ratio $R^{J}_{\sigma}$ from the source $S_1$ of Eq.~(\ref{ratio}) versus $t/a$  
for a source-sink
separation 0.91~fm shifted by a time slice (triangles)
and 1.14~fm (squares)
for the four smallest non-zero $\vec{q}^2$ values. The fit range is also shown along with the fitted lines and the corresponding error bands. The behavior is the same for both, but the error reduction is better in the former, which is what we therefore utilize in the calculations.
}
\label{fig:plateaus3}%
\end{center}
\end{figure}

Since the source-sink separation is fixed in this method
 it is crucial to suppress the excited baryon states as much as
possible. This is achieved by employing  gauge invariant Gaussian smearing on the local quark fields with APE-smeared gauge fields
and parameters that have been carefully optimized for the nucleon state. 
 For the coarse lattice, we show in Fig.~\ref{fig:plateaus3} a comparison
of results obtained with a sink-source separation of 0.91~fm and 1.14~fm.
As can be seen,
extending the source-sink separation to 1.14~fm, the plateau values for the dominant magnetic dipole form factor
$G_{M1}$, which are the most accurate, are consistent with a time-separation
of 0.91~fm. Since the larger time separation introduces a doubling in the statistical noise, for the accuracy needed 
in this study, we opt to use the smaller sink-source separation in time.
For the fine lattice, we take a sink-source separation of $\Delta T = 12a$ 
corresponding to 1.01~fm, consistent with our findings using the
coarse lattice.
%

%
%\begin{figure}[htb]
%\centering
%\subfigure[][]{\label{fig:plateaus3}%
%\includegraphics[width=0.48\linewidth,height=0.48\linewidth]{plateaus3} }
%\hspace{5pt}
%\subfigure[][]{\label{fig:GM1coherentsinks}%
%\includegraphics[width=0.48\linewidth,height=0.48\linewidth]{GM1_coherent_sinks}}
%\caption{\subref{fig:plateaus3} 
%{\bf Dina: These hardly show any plateaus. Please fit and plot
%the value with its error band.
%{\textcolor{red}{ \bf Yiannis: This plot was provided by AT and it shows just the coarse %lattice ratios. It compares the errors between time separations
%$t_f/a=8$ and $t_f/a=10$, I think this serves  this purpose.}}} 
%The ratio $R_{\sigma}$ from the source $S_1$ of Eq.~(\ref{ratio}) versus $t/a$  
%for a source-sink
%separation 0.91~fm shifted by a time slice (blue triangles)
%and 1.14~fm (red circles)
%for the four smaller non-zero $\vec{q}^2 $. The behavior is the same for both, but the error %reduction is better in the former, which we utilize in the calculations.
%\subref{fig:GM1coherentsinks}  Dependence of the jackknife error for $G_{M1}(Q^2)$ on the %coherent sink bin sizes. This test shows that there is no
%problem with  auto-correlations in the coherent sink method applied in this study. 
%}
%\label{fig:timesep-GM1sinks}
%\end{figure}
%
 
In order to 
improve accuracy, a goal that is particularly crucial 
for the extraction of the sub-dominant electromagnetic form factors, we
employ a new method first implemented in the
study of the nucleon form factors~\cite{Syritsyn:2009mx} and referred to
 as the {\it coherent sink 
technique}.
 The method consists of creating four sets of forward 
propagators for each configuration at source positions
separated in time by one-quarter of the total temporal size. 
Namely, for the coarse DWF lattice, $N_L=24,$ we have forward propagators generated 
with sources positioned at: 
\[ \bigg \{(\vec{0},0), \ (\frac{\vec{L}}{2},16a), \ (\frac{\vec{L}}{4},32a),\ (\frac{\vec{3L}}{4},48a)\bigg \} ,\]
and for the fine DWF lattice, $N_L=32,$ placed at:
$$ \bigg \{(\vec{0},10a), \ (\frac{\vec{L}}{2},26a), \ (\vec{0},42a),\ (\frac{\vec{L}}{2},58a)\bigg \},$$
or
$$ \bigg \{(\frac{\vec{L}}{4},10a), \ (\frac{3\vec{L}}{4},26a), \ (\frac{\vec{L}}{4},42a),\ (\frac{3\vec{L}}{4},58a)\bigg \}.$$

\begin{figure}[htb]
\begin{center}
\hspace{-0.2cm}\includegraphics[width=0.65\linewidth,height=0.55\linewidth]{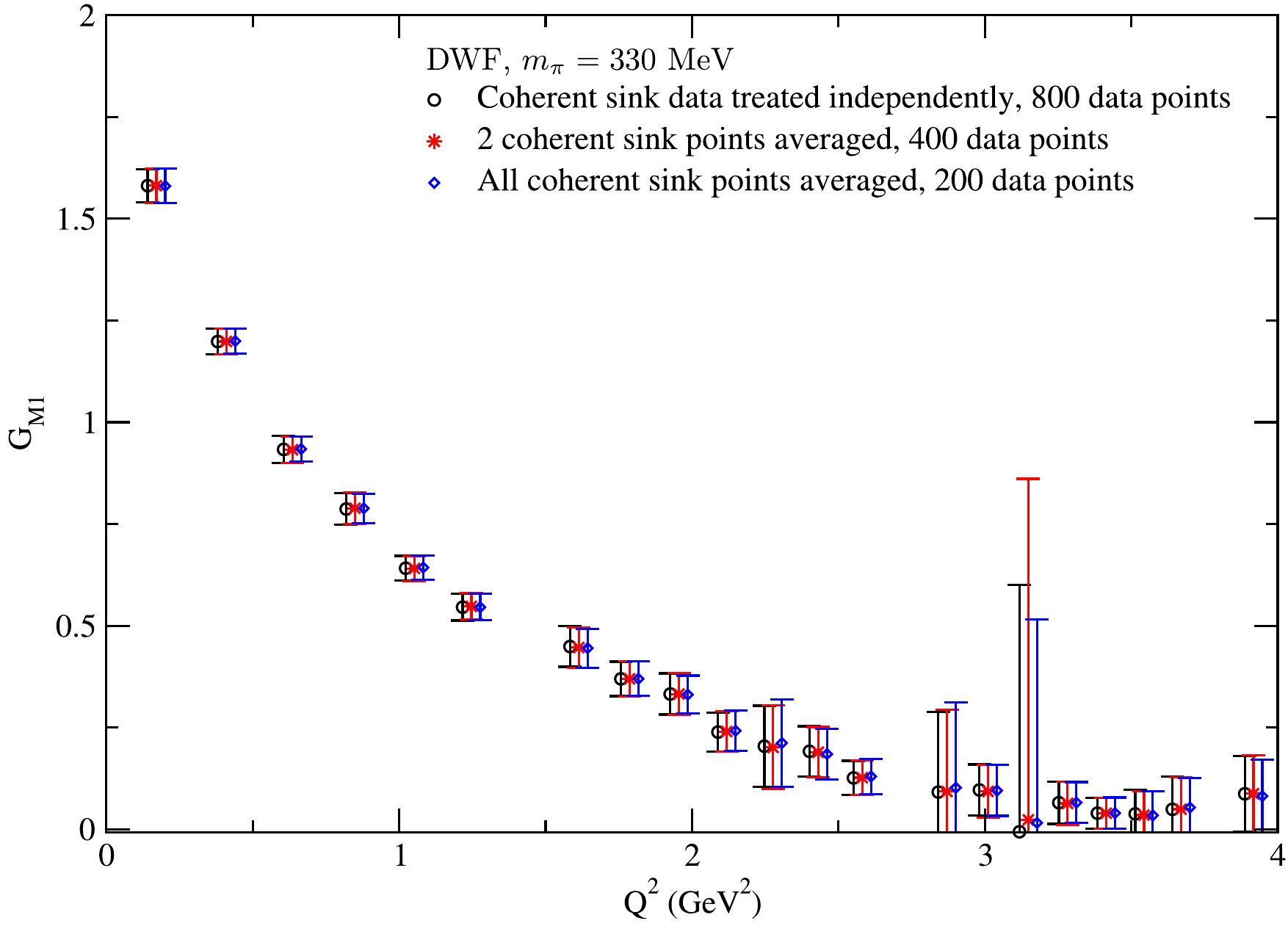} 
\vspace*{-0.5cm}
\caption{
Dependence of the jackknife error for $G_{M1}(Q^2)$ on the coherent sink bin sizes. This test shows that there is no
problem with cross-correlations in the coherent sink method applied in this study.
}
\label{fig:timesep-GM1sinks}%
\end{center}
\end{figure}

From each source $(\vec{x}_i,T_i)$,
a zero-momentum projected $\Delta$ source is constructed at $T_0$ slices away, i.e. at
$(\vec{x}_i,T_i+T_0).$ For the coarse DWF lattice $T_0/a=8,$ while for the fine DWF lattice $T_0/a=12$. Then 
a \emph{single} coherent backward propagator is calculated in the
simultaneous presence of all four sources. The cross terms
that arise vanish due to gauge invariance when averaged over the ensemble.
The forward propagators  have already been computed by the LHPC collaboration~\cite{Syritsyn:2009mx} and therefore
we effectively obtain four measurements at the cost of one sequential  inversion. 
This assumes large enough time-separation between the four sources to suppress contamination among them. A question that arises 
is whether or not there exist statistically important correlations among these four measurements. In Fig.~\ref{fig:timesep-GM1sinks} 
we show the dependence of the jackknife error on the magnetic dipole 
$G_{M1}$ for different coherent sink bin sizes. As can be seen, the jackknife errors using one sequential inversion for each are the same as combining all four
in single inversion.  This is a direct  verification that cross-correlations between the different sinks 
are absent or negligible.

Finally, the full set of lattice data obtained at a given $Q^2$ value is analyzed simultaneously
by a global $\chi^2$ minimization using the singular value decomposition of an
overconstrained linear system~\cite{Alexandrou:2007xj,Hagler:2003jd}. Generically, this consists of setting up the  following linear over-complete system of equations
\begin{align}\label{svd1}
P({\bf q};\mu)= D({\bf q};\mu)\cdot F(Q^2),
\end{align} 
where $P({\bf q};\mu)$ represent the lattice measurements of the appropriately
defined ratios of Eq.~\ref{ratio}, each one with its associated  statistical weight $w_{k}$. The column vector $F(Q^2)$ contains the number $M$ of form factors to be extracted. If we let $N$ represent the number of momentum vectors $\mathbf{q}$ and current directions $\mu$ that contribute to a specific value of $Q^2,$ then  $D({\bf q};\mu)$ 
is a matrix structure of the form $N\times M$ which depends on kinematical form factors obtained from the trace algebra on the employed matrix element. The form factors, at the specific $Q^2$ value, are then extracted from the minimization of the total $\chi^2$:
\begin{align}\label{chisq}
\chi^2=\sum_{k=1}^{N} \Biggl(\frac{\sum_{j=1}^2 D_{kj}F_j-P_k}{w_k}\Biggr)^2, 
\end{align}
by applying the singular value decomposition on the $N\times M,$  $D({\bf q};\mu)$ matrix. All the errors on the lattice 
measurements as well as the errors on the form factors are determined from the jackknife procedure.
\section{\label{sec:EM-ffs} Electromagnetic N--to--$\Delta$ Transition form factors}
\subsection{\label{sec:EM_mat_el} The electromagnetic matrix element}

The electromagnetic transition matrix element
\be
  \langle\Delta(p',s')\vert j_\mu \vert N(p,s)\rangle =
  i \,\sqrt{\frac{2}{3}} \; \biggl(\frac{m_{\Delta}\; m_N}
{E_{\Delta}({\bf p}^\prime)\;E_N({\bf p})}\biggr)^{1/2} \bar{u}_\sigma (p',s')
  {\cal{O}}_{\sigma\mu} u(p,s) 
\ee
 is decomposed in terms of 
three multipole form factors:
$$  {\cal O}_{\sigma\mu} = G_{M1} (q^2) K_{\sigma\mu}^{M1} + G_{E2}(q^2)
  K_{\sigma\mu}^{E2} + G_{C2} (q^2) K_{\sigma\mu}^{C2} 
$$
where the kinematical factors in  Euclidean space are given by
\begin{align}
  \label{eq:lorentz-op-prefac}
  K_{\sigma\mu}^{M1} &= - \frac{3}{(m_\Delta+m_N)^2+Q^2}
  \frac{m_\Delta+m_N}{2m_N}\,\imag\,
  \varepsilon_{\sigma\mu\alpha\beta}\, p^\alpha 
  {p'}^\beta\,, \nonumber \\
  K_{\sigma\mu}^{E2} &=  -K_{\sigma\mu}^{M1} + 6\, \Omega^{-1}(Q^2)
  \frac{m_\Delta+m_N}{2m_N} 2\,\imag\, \gamma_5\,
  \varepsilon_{\sigma\lambda\alpha\beta}\, p^\alpha {p'}^\beta
  \varepsilon_{\mu}^{\phantom{\mu}\lambda\gamma\delta}\, p_\gamma
  {p'}_\delta\,, \nonumber\\
  K_{\sigma\mu}^{C2} &= -6\, \Omega^{-1}(q^2)
  \frac{m_\Delta+m_N}{2m_N}\, \imag\, \gamma_5\, q_\sigma \left( q^2
    (p+p')_\mu - q \cdot (p+p') q_\mu \right).
\end{align}
The $p(s)$ and $p'(s')$ denote initial and final momenta (spins),  
$q^2\equiv(p'-p)^2$,  and
$ u_\sigma (p',s')$ is a Rarita-Schwinger vector-spinor. We also define $\Omega(Q^2) =  \left[(m_\Delta+m_N)^2+Q^2\right] \left [(m_\Delta-m_N)^2+Q^2\right],$ 
with (${\bf Q}={\bf q}$, $Q_4=iq^0$), so the lattice momentum transfer gives $Q^2=-q^2$.

In this work we present results for the dominant magnetic dipole 
form factor $G_{M1}(q^2)$ as well as the sub-dominant electric $G_{E2}(q^2)$ and Coulomb quadrupole $G_{C2}(q^2)$ form factors. 
Note that these are all scalar functions depending on the momentum transfer $q^2=-Q^2,$ whereas on the lattice only the space-like $q^2$ are accessible, thus $Q^2>0.$

\subsection{\label{sec:GM1} The magnetic dipole form factor}

The magnetic dipole form factor is directly evaluated from the
optimized linear combination $S^V_1$ with  the vector current $V_\mu(x)$ insertion.
In the large Euclidean time separation limit with the $\Delta$ produced at zero momentum we obtain,
\be
\label{S1} 
S^V_1({\bf q};V_\mu)= 
 i A \biggl\{ (p_2-p_3)\delta_{1,\mu} 
 + (p_3-p_1)\delta_{2,\mu} + (p_1-p_2)\delta_{3,\mu} \biggr\}
G_{M1}(Q^2) \quad. 
\ee
The vector index $\mu$ takes spatial values, $\mu =1,2,3$ and $A$ is a kinematical constant,
\be
A = \sqrt{\frac{2}{3}} \frac{m_\Delta + m_N}{4m_N E_N} \sqrt{\frac{E_N}{E_N+m_N}} ~.
\label{A}
\ee
The local vector current of Eq.~(\ref{veccurr}) is not conserved by the lattice action and the renormalization constant $Z_V$, given in 
Table~\ref{Table:params_DWF_hybrid}, is used to renormalize the current. $Z_V$ is determined from charge conservation that dictates that the electric 
 nucleon form factor is one at $Q^2=0$, namely  $Z_V = 1/G_E(0)=1/F_1(0)$ where $F_1$ is the Dirac form factor.

\begin{figure}[htb]
\begin{center}
\hspace{-0.2cm}\includegraphics[width=0.65\linewidth,height=0.65\linewidth]{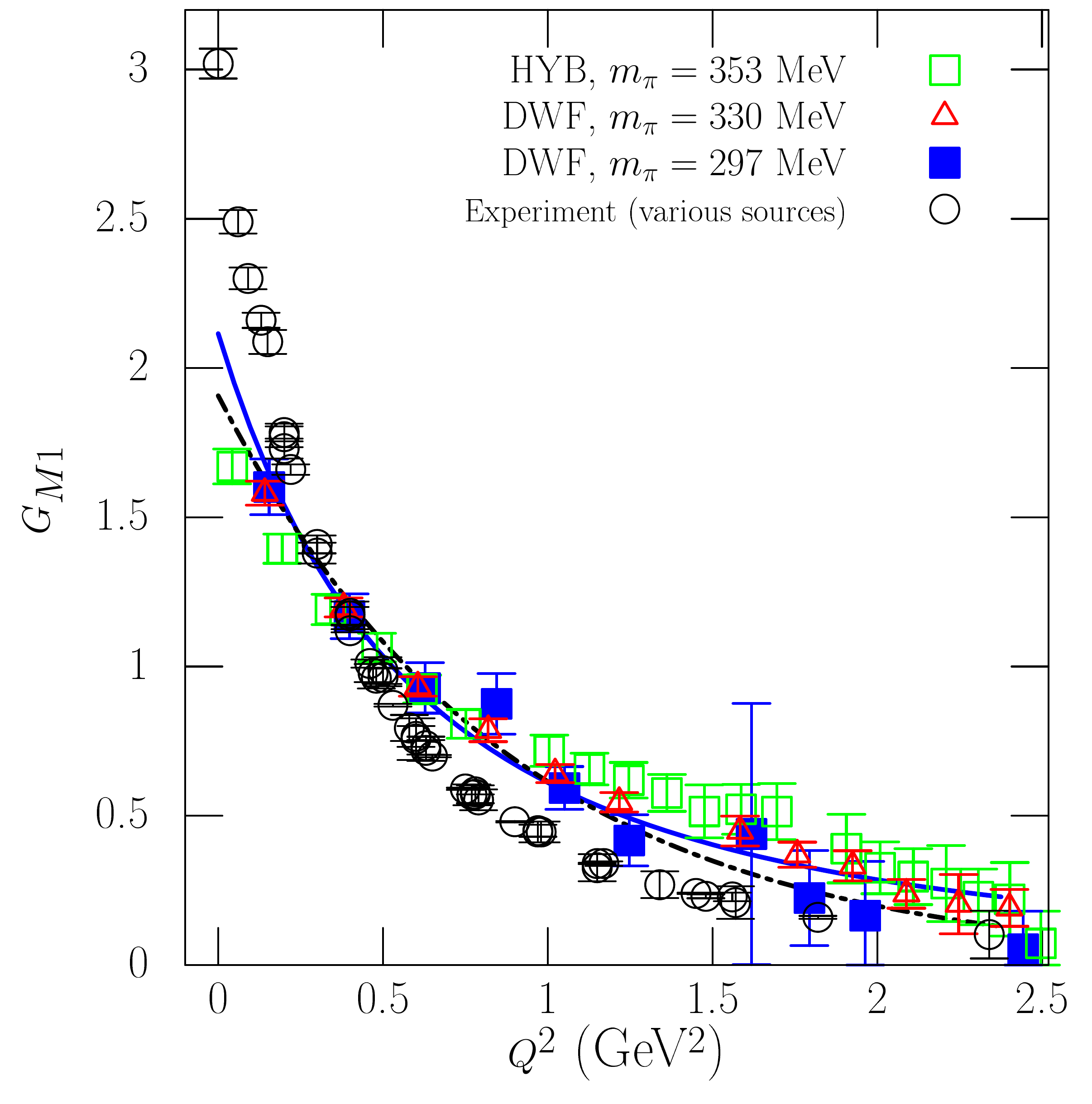} 
\vspace*{-0.5cm}
\caption{
The magnetic dipole $G_{M1}(Q^2)$ using DWF 
fermions (both coarse and fine lattices) and using the hybrid action. 
The circles show the experimental results.
The solid blue (dashed black) line is a fit to 
dipole (exponential) form for the fine DWF lattice.}.
\label{fig:GM1fit}
\end{center}
\end{figure}

In Fig.~\ref{fig:GM1fit} we show the DWF results for the magnetic dipole form factor $G_{M1}$  at $m_{\pi}=330$~MeV on the coarse lattice and 
at $m_{\pi}=297$~MeV on the fine lattice
as a function of the momentum transfer $Q^2$.
These are compared  with our previous results obtained with a hybrid action approach that uses
Asqtad improved staggered fermions generated by the MILC collaboration~\cite{Bernard:2001MILC} and
domain wall valence quarks~\cite{Alexandrou:2007dt}. 
The experimentally available data (for more
details see Ref.~\cite{Alexandrou:2007dt}) are also shown in Fig.~\ref{fig:GM1fit}
showing a  discrepancy between lattice results and experiment.
Although there is a small decrease in the value of $G_{M1}$
 at high $Q^2$ bringing lattice data closer to experiment, the slope at
small $Q^2$ is still smaller than in experiment.  The second observation is that
although the hybrid calculation used a lattice spacing about 50\% larger than
the fine DWF lattice, these data show no significant finite $a$-effects. 
Fits to a dipole form,  $g_0/(1+Q^2/m_0^2)^2$, as well as to an 
exponential form 
$\tilde{g_0} 
\;\mathrm{exp}(-Q^2/\tilde{m}_{0}^2)$ are shown for
the fine DWF lattice. As can be seen they both
provide a good description of the lattice results.  A list of the fit parameters for all sets is provided in Table~\ref{Table:GM1fitparams}.
%\noindent
%

The discrepancy between experiment and lattice results
is clearly reflected in the value of the  dipole mass of $m_0 = 0.78$~GeV
obtained by performing a dipole fit to the experimental data
as compared to the  $m_0$ values obtained from  the lattice results  listed in Table~\ref{Table:GM1fitparams}.
The steeper rise of the experimental results on $G_{M1}$ as a function of $Q^2$ near the origin
is  indicative of the onset of strong
chiral quark effects, or equivalently, the lack of strong pion cloud from the
still heavy pion mass lattice ensembles that are utilized. Similar 
behavior has also been observed in the nucleon electromagnetic 
form factors studies~\cite{Alexandrou:2009ng,Yamazaki:2009zq,Syritsyn:2009mx}. 
The N to $\Delta$
 transition is particularly clean since there is
no ambiguity regarding disconnected contributions and thus the flatter 
dependence observed in the N-to-$\Delta$ electromagnetic form factor  must be of different origin.
Large pion cloud effects would have to set in as we lower the pion mass in order to explain the experimental curve. 
Such effects have been shown to arise in chiral expansions~\cite{Pascalutsa:2005} and it is thus  interesting to repeat 
the calculation for $m_\pi<250$~MeV where they are expected to become more
pronounced. 
%
%GM1--fit params
%
\begin{widetext}
\begin{center}
\begin{table}[h]
\begin{tabular}{ccccc}
 \hline \hline  
$m_\pi$~[GeV] & $g_0$ &  $m_0$ [GeV] & $\tilde{g}_0$ &  $\tilde{m}_0$ [GeV] \\ 
\hline
\multicolumn{4}{l}
{coarse $N_F=2+1$ DWF}\\
0.329(1) & 1.937(65)   & 1.171(44) & 1.737(53) & 1.025(32) \\
\hline
\multicolumn{4}{l}
{fine $N_F=2+1$ DWF}\\
0.297(5) & 2.115(161)& 1.078(79) & 1.907(127) & 0.939(55)\\
\hline
\multicolumn{4}{l}
{HYBRID}\\
0.353(2) & 3.263(64)& 1.305(27) & 3.05935(61) & 1.097(21)\\
\hline
\multicolumn{4}{l}
{Experiment- (various sources) }\\
-- & 3.266(40)& 0.745(4) & 2.202(67) & 0.776(10)\\
\hline\hline
\end{tabular}
\caption{The fit parameters for the magnetic dipole form factor obtained for both a dipole and an exponential fit form with fit parameters ($g_0,\ m_0$) and ($\tilde{g}_0,\ \tilde{m}_0$), respectively.}
\label{Table:GM1fitparams}
\end{table}
\end{center}
\end{widetext}
\subsection{\label{sec:subGE2} The electric quadrupole form factor--$G_{E2}$}   

The sub-dominant electromagnetic quadrupole form factors $G_{E2}$ and $G_{C2}$ are  extracted from the 
optimized sources $S^V_2$ and $S^V_3$. 
The relevant expressions  for a static $\Delta$ final state are~\cite{Alexandrou:2007dt}:
\begin{align} 
\label{S2} 
S^V_2({\bf q};\mu)=
&-3 A \Biggl\{  \bigl( (p_2+p_3)\delta_{1,\mu}
 + (p_3+p_1)\delta_{2,\mu} +
(p_1+p_2)\delta_{3,\mu} \bigr ) {\cal G}_{E2}(Q^2) \nonumber \\ 
&- 2 \frac{p_\mu}{{\bf p}^2}\bigl( p_1 p_2 + p_1 p_3 +
p_2 p_3 \bigr) 
\left[ {\cal G}_{E2}(Q^2) + \frac{E_N-m_\Delta}{2 m_\Delta} 
{\cal G}_{C2}(Q^2)
\right] \Biggr\} \; , 
\end{align}
 for  the spatial current directions $\mu = 1,2,3 $. For the temporal current direction $\mu =4,$ we have 
\be \label{S2b} 
S^V_2({\bf q};\mu = 4)= 
\frac{-i\; 6\;B}{{\bf p}^2} (p_1 p_2 + p_1 p_3 + p_2 p_3)
{\cal G}_{C2}(Q^2)\;,
\ee 
where $B$ is given by $B = \frac{{\bf p}^2}{2m_\Delta} A$, and $A$ is the constant provided in Eq.~\ref{A}.

Notice that the above combination, if used alone, will not allow for the extraction of $G_{C2}$ at the lowest photon 
momentum $\mathbf{q}=(1,0,0)\frac{2\pi}{aL}.$ Since chiral effects are stronger at low $Q^2$ values and experiments are targeted
in that regime, we utilize the optimal linear three-function combination $S^V_3$ in order to obtain $G_{C2}$ also at the 
lowest $Q^2$ point allowed on the lattice. The corresponding expressions are
\begin{align} \label{S3}
S^V_3({\bf q};\mu)=& 
-\frac{3\; A}{2}\; p_\mu \; \Biggl[ 
3\; \biggl( \delta_{\mu,3}-\frac{p_3^2}{{\bf p}^2} \biggr)\; {\cal G}_{E2}(Q^2) 
+\frac{E_N-m_\Delta}{2 m_\Delta} \; \biggl(1 - 3\;\frac{p_3^2}{{\bf p}^2} \biggr) \; {\cal G}_{C2}(Q^2) 
\Biggr]
\end{align}
for $\mu = 1,2,3$ and for the temporal component
\beq \label{S3b}
S^V_3({\bf q};\mu = 4)= 
\frac{3\; i\;B}{2} \biggl( 1-3\; \frac{p_3^2}{{\bf p}^2}  \biggr) \;
{\cal G}_{C2}(Q^2), 
\eeq
which is directly proportional to $G_{C2}(Q^2).$
Data obtained from both $S^V_2$ and $S^V_3$ are simultaneously fitted in the overconstrained analysis in order to 
extract the momentum dependence of $G_{E2}$ and $G_{C2}$ as accurately as possible.

\begin{figure}[htb]
\centering
\subfigure[][]{\label{fig:GE2ff}%
\includegraphics[width=0.48\linewidth,height=0.48\linewidth]{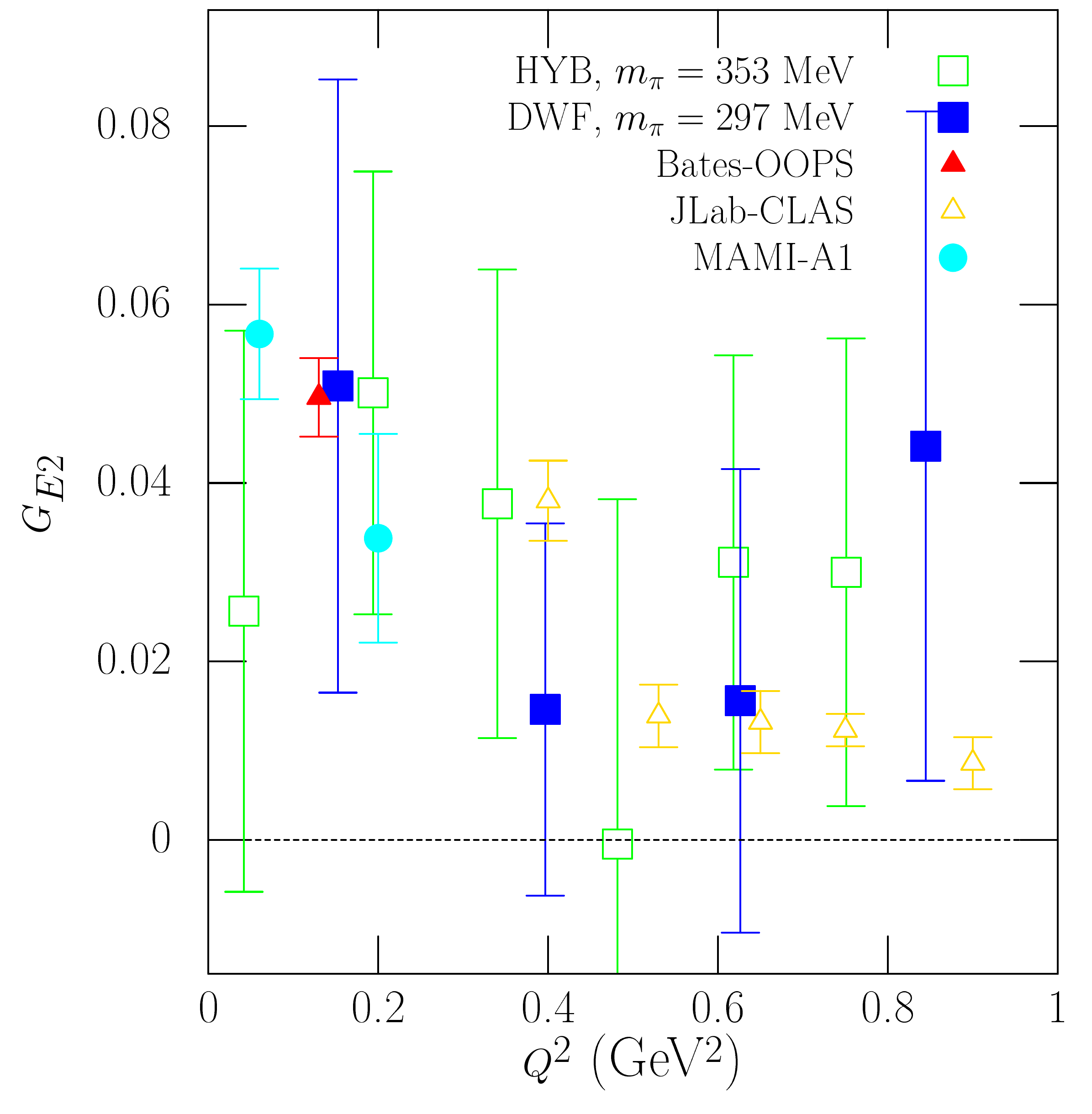} }
\hspace{-3pt}
\subfigure[][]{\label{fig:EMR}%
\includegraphics[width=0.48\linewidth,height=0.48\linewidth]{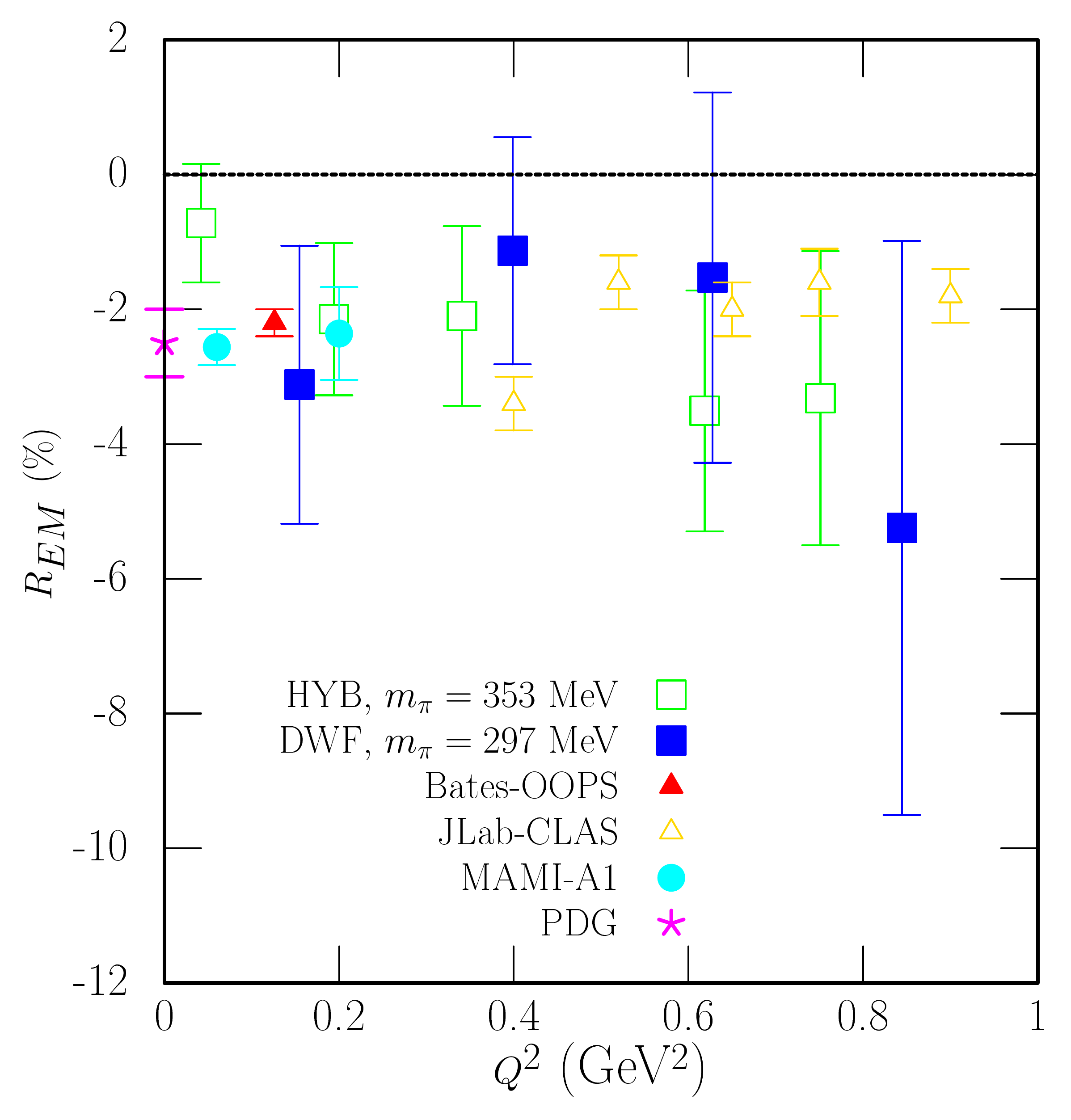}}
\caption{In \subref{fig:GE2ff}  the result of the electric quadrupole form factor $G_{E2}(Q^2)$ extracted from the fine DWF lattice measurements is shown. The results obtained from the hybrid action~\cite{Alexandrou:2007dt}, as well as the experimentally extracted results from Bates~\cite{Mertz:2001,Sparveris:2002fh,Sparveris:2004jn}, Jlab~\cite{Joo:2001} and MAMI~\cite{Stave:2006ea,Sparveris:2006uk} are also plotted for comparison.
In \subref{fig:EMR}  the corresponding $R_{EM}$ evaluated in the rest frame of the $\Delta$ baryon ($\mathbf{p^{\prime}}=0$) is depicted for
the fine DWF lattice as well as
 for the hybrid action~\cite{Alexandrou:2007dt}. The experimentally available results from~\cite{Mertz:2001,Sparveris:2002fh,Sparveris:2004jn,Joo:2001, Stave:2006ea,Sparveris:2006uk} are also shown.
}\label{fig:GE2-EMR}
\end{figure}
In Fig.~\ref{fig:GE2ff} we plot the values of the  electric quadrupole form factor $G_{E2}$ for a range of values of 
$Q^2 < 1$~GeV$^2$, in the case of the fine DWF lattice. These results are compared to the results obtained from the mixed action~\cite{Alexandrou:2007dt}. We also mention here that in the case of the coarse DWF lattice 
the statistical noise on the $G_{E2}$ and $G_{C2}$ values is  large, so
a  zero value can therefore not be excluded. 
The phenomenologically interesting ratio $R_{EM}$ is defined as
\begin{align} \label{R_EM}
R_{EM}=-\frac{G_{E2}(Q^2)}{G_{M1}(Q^2)},
\end{align}
and has been used traditionally as a signal of deviation from spherical symmetry in the nucleon-$\Delta$ system. Early quark
models as well as models of the proton wave function based on relativistic quarks including two-body exchange currents agree that
a small $R_{EM}$ value in the $-1\sim 2\% $ regime should appear.
 The experimental values included in Fig.~\ref{fig:GE2ff} show 
practically no dependence on $Q^2$. The same is true for the lattice data and in fact a good consistency with the 
experiment is evident.
The approach to the physical point can be predicted
in chiral effective theory~\cite{Pascalutsa:2005} where a non-monotonic
dependence on the pion mass is expected with a minimum at 200 MeV. It is a significant challenge for the lattice to provide accurate results in the future in this regime in order to crosscheck 
 the pion dynamics.

\subsection{\label{sec:subGC2} The Coulomb quadrupole form factor--$G_{C2}$}   

\begin{figure}[htb]
\centering
\subfigure[][]{\label{fig:GC2ff}%
\includegraphics[width=0.48\linewidth,height=0.48\linewidth]{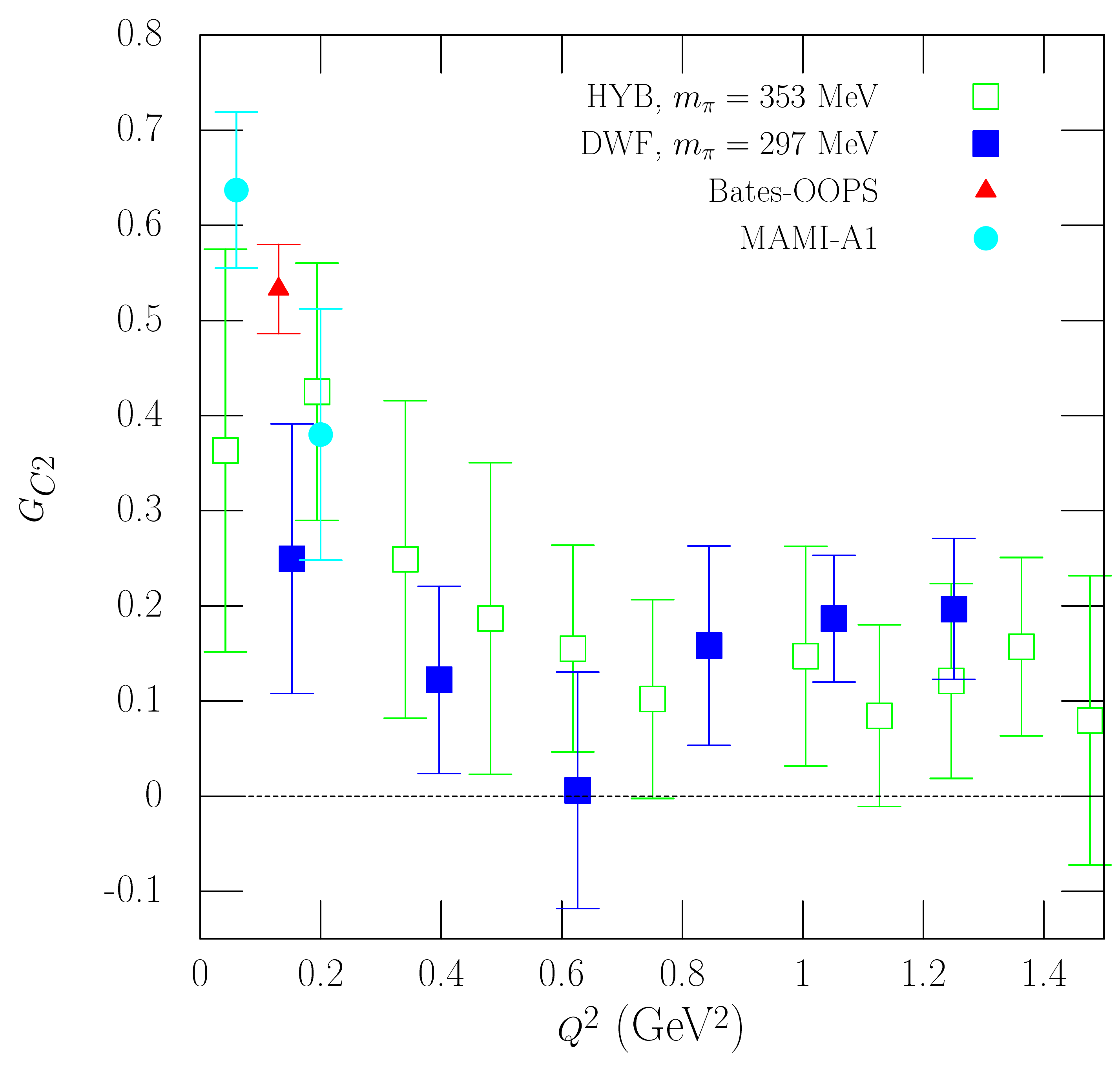} }
\hspace{-3pt}
\subfigure[][]{\label{fig:CMR}%
\includegraphics[width=0.48\linewidth,height=0.48\linewidth]{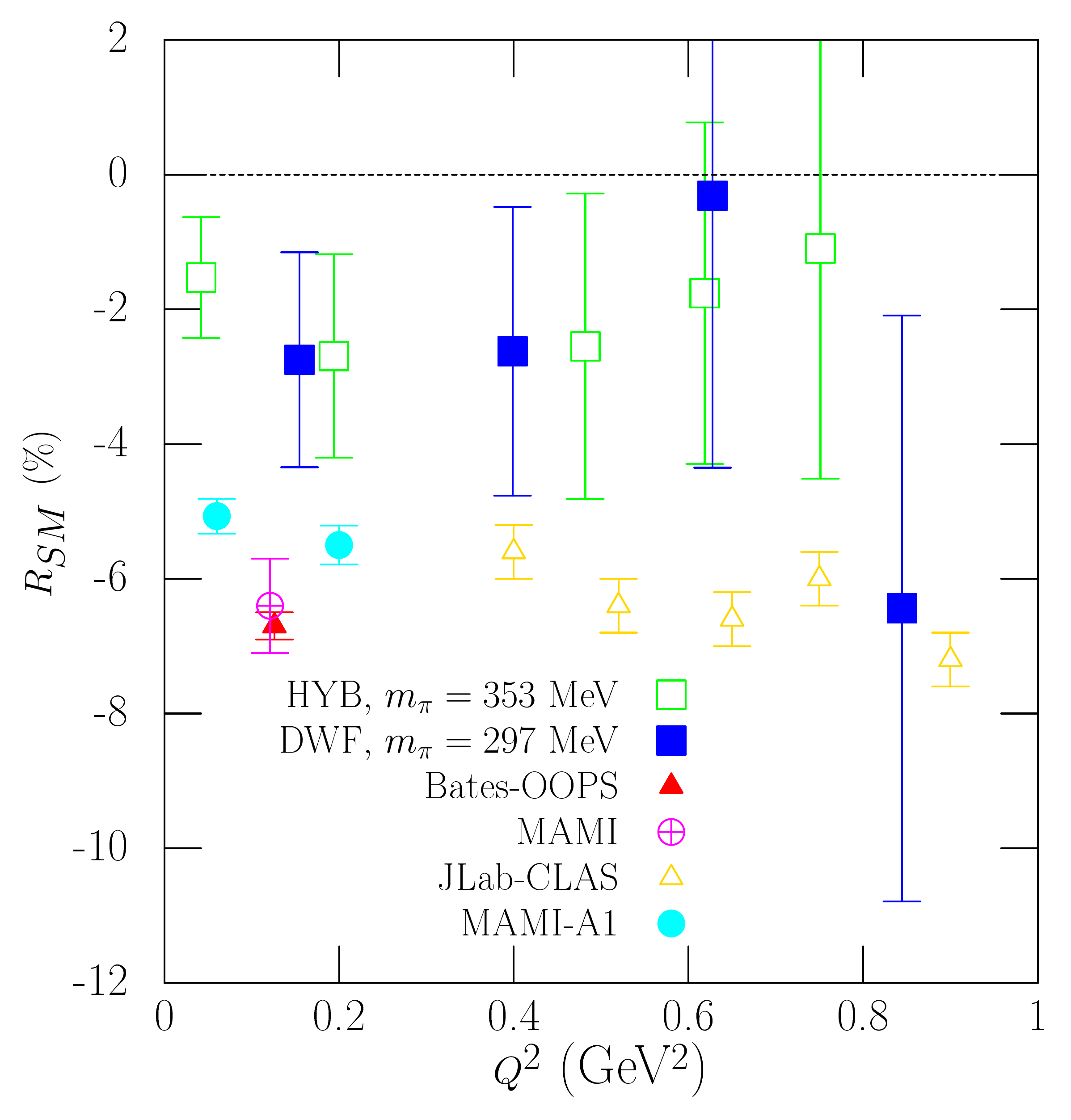}}
\caption{In plot \subref{fig:GC2ff} we  show the Coulomb quadrupole form factor $G_{C2}(Q^2)$ extracted from the fine DWF 
lattice measurements. Along with it we provide also the result from the hybrid action approach~\cite{Alexandrou:2007dt}.
Plot \subref{fig:CMR}  depicts the corresponding $R_{SM}$ evaluated in the rest frame of the $\Delta$ baryon. Non-zero values
are confirmed, for the lowest $Q^2$ values accessible on the lattices. 
We also show results using  the hybrid action taken from Ref.~\cite{Alexandrou:2007dt}. Experimental results are also included using the same notation as those
in Fig.~\ref{fig:GE2-EMR}.
}\label{fig:GC2-CMR}
\end{figure}

As mentioned in the previous section, the Coulomb quadrupole form factor is computed with the help of Eqs.~(\ref{S2})-(\ref{S3b}). 
In the case of $G_{C2}$, Fig.~\ref{fig:GC2ff} shows the results from the fine DWF lattice for values of $Q^2<1.5$~GeV$^2$. 
The values of $G_{C2}(Q^2)$ are positive and consistent with previous results obtained using the mixed action~\cite{Alexandrou:2007dt}, and are also shown on the same figure. 
The experimentally measured ratio of Coulomb quadrupole to magnetic dipole
form factor known also as CMR is defined by
\begin{align} \label{R_SM}
R_{SM}=-\frac{|\mathbf{q}|}{2 m_{\Delta}}\frac{G_{C2}(Q^2)}{G_{M1}(Q^2)},
\end{align}
in the frame where the $\Delta$ is produced at rest. Lattice results on the $R_{SM}$ ratio are shown in Fig.~\ref{fig:CMR}
where $m_{\Delta}$ in Eq.~(\ref{R_SM}) is set to the physical mass.
Known values of $R_{SM}$ from various experiments are included in Fig.~\ref{fig:CMR} and as with $R_{EM}$ show almost no
dependence on the momentum transfer. This is also the feature shown by the two lattice ensembles, the fine DWF at 297~MeV and
the hybrid scheme  at 353~MeV which are in very good agreement with each other. Despite the large statistical errors which 
escort the lattice values, they disagree with the experiment.
Chiral effective theory predicts a monotonic decrease of this
ratio as the pion mass approaches the chiral limit, which is different
from the dependence of $R_{EM}$. The onset of large pion effects are expected 
below 300 MeV pions.

The overall conclusion is that QCD  confirms non-zero quadrupole
amplitudes pointing to the existence of the deformation in the $N-\Delta$ system, as coded in the
EMR and CMR ratios.
However, quantitative agreement with experiment has to await
simulations at lighter pions masses, expected to 
become available in the next couple of years. Statistical accuracy at these
light pions masses in particular for the sub-dominant form factors
is an issue that has to be addressed. The use of the coherent source technique
as employed here is a way to increase statistical accuracy.
\section{\label{sec:AX-ffs} Axial N to $\Delta$ transition form factors and the
Goldberger-Treiman relation}
\subsection{\label{sec:EWeakmatrixEl} The Electro-weak and Pseudo-scalar transition matrix element}

The   nucleon to $\Delta$ matrix element of the axial vector current
is parameterized in terms of 
four dimensionless form factors. In the Adler parameterization~\cite{Alder:1972}
it is written as follows
\beq
\langle \Delta(p^{\prime},s^\prime)|A^3_{\mu}|N(p,s)\rangle &=& i\sqrt{\frac{2}{3}} 
\left(\frac{m_\Delta m_N}{E_\Delta({\bf p}^\prime) E_N({\bf p})}\right)^{1/2}
\bar{u}_{\Delta^+}^\lambda(p^\prime,s^\prime)\nonumber \\
&\>&\hspace*{-5cm}
\biggl[\left (\frac{C^A_3(q^2)}{m_N}\gamma^\nu + \frac{C^A_4(q^2)}{m^2_N}p{^{\prime \nu}}\right)  
\left(g_{\lambda\mu}g_{\rho\nu}-g_{\lambda\rho}g_{\mu\nu}\right)q^\rho
+C^A_5(q^2) g_{\lambda\mu} +\frac{C^A_6(q^2)}{m^2_N} q_\lambda q_\mu \biggr]
u_P(p,s)
\label{NDaxial}
\eeq
with the axial current given in Eq.~(\ref{currents}). 
%\beq
%\label{AxialCur}
%A_{\mu}^a(x)&= \bar{\psi}(x)\gamma_\mu \gamma_5\frac{\tau^a}{2}\psi(x) ~,
%\eeq
%where $\tau^a$ ($a=1,2,3$) are the three Pauli-matrices acting in flavor space
%and $\psi$ is the  isospin doublet quark field.

The form factors
$C^A_3(q^2)$ and $C^A_4(q^2)$ belong to the transverse part of the axial 
current and are both suppressed~\cite{Alexandrou:2006mc} relative to the 
longitudinal form factors $C^A_5(q^2)$ and   $C^A_6(q^2)$, which are  the dominant ones and are the ones considered in this work.

Likewise, the pseudo-scalar transition form factor 
$G_{\pi N\Delta}(q^2)$, is defined via
%\small
\be 
 2m_q\langle \Delta(p^\prime,s^\prime)|P^3|N(p,s)\rangle = i\sqrt{\frac{2}{3}}
\left(\frac{m_\Delta m_N}{E_\Delta({\bf p}^\prime) E_N({\bf p})}\right)^{1/2}
\frac{f_\pi m_\pi^2 \>G_{\pi N\Delta}(q^2)}
{m_\pi^2-q^2}
\bar{u}_{\Delta^+}^\nu(p^\prime,s^\prime)\frac{q_\nu}{2m_N} u_P(p,s)
\label{gpiND} 
\ee
%\normalsize
where the normalization of the RHS of~(\ref{gpiND}) is chosen such that 
$G_{\pi N\Delta}(q^2)$ reproduces the phenomenological coupling of the $\pi-N-\Delta$ vertex
in the strong interaction Lagrangian,
\be
{\cal L}_{\pi N \Delta} = \frac{g_{\pi N\Delta}}{2 m_N} \bar{\Delta}_\mu \partial_\mu \vec{\pi} \cdot \vec{\tau}~ N + {\rm h.c.}
\ee 
and the pseudo-scalar density is  defined in Eq.~(\ref{currents}).
%\beq
%P^a(x)&= \bar{\psi}(x)\gamma_5 \frac{\tau^a}{2}\psi(x) ~, 
%\label{PseudCur}
%\eeq
In the SU(2) symmetric limit with $m_q$ denoting the up/down mass, the pseudo-scalar density is related  
to the divergence of the axial-vector current through the
axial Ward-Takahashi identity (AWI)
\begin{align} \label{AWI}
 \partial^\mu A_\mu^a = 2 m_q P^a ~.
\end{align}
Taking matrix elements of the above identity between N and $\Delta$ states
leads to the non-diagonal Goldberger-Treiman (GT) relation
\beq
 C_5^A(q^2)+\frac{q^2}{m_N^2} C_6^A(q^2) = 
\frac{1}{2m_N}\frac{G_{\pi N \Delta}(q^2)f_\pi m_\pi^2}{m_\pi^2-q^2} \quad.
\label{GTR_ND}
\eeq
On the other hand, flavor symmetry in the hadronic world is expressed through the 
partially-conserved axial vector current (PCAC) hypothesis 
\begin{align}\label{PCAC}
\partial^\mu A_\mu^a=f_\pi m_\pi^2 \pi^a
\end{align}
which relates the pseudo-scalar current to the pion field operator and the pion decay constant $f_\pi$ which is here is taken to be  $92$~MeV.
From Eqs.~(\ref{AWI}) and (\ref{PCAC}) the pion field $\pi^a$ is related to the pseudo-scalar density via
\beq
\pi^a = \frac{2 m_q P^a}{f_\pi m_\pi^2}.
\eeq
%{\bf Dina: I am not sure I understand the following so I commented out}
%and therefore provides the connection to the 
%phenomenological $\pi N\Delta$ strong coupling, $g_{\pi N\Delta}$,
%strictly valid at the chiral limit of QCD,
%\be
%g_{\pi N\Delta} = G_{\pi N\Delta}(q^2 \simeq m_\pi^2 \simeq 0) ~.
%\ee
Assuming pion pole dominance we can
relate the form factor $C_6^A$ to $G_{\pi N\Delta}$
through:
\beq 
\frac{1}{m_N}C_6^A(q^2)&\sim&\frac{1}{2}\frac{G_{\pi N\Delta}(q^2) f_\pi}
{m_\pi^2-q^2}
\label{GP}
\eeq
Then, substituting Eq.~(\ref{GP}) in Eq.~(\ref{GTR_ND}), we obtain the simplified
Goldberger-Treiman (GT) relation
\beq
G_{\pi N \Delta}(q^2)\>f_\pi &=& 2m_N C_5^A(q^2)
\label{GTR}
\eeq
in an  analogous fashion to the well known GT relation which holds in the
nucleon sector studied on the lattice in Ref.~\cite{Alexandrou:2007xj}. Pion pole dominance therefore fixes completely the ratio 
$C_6^A(q^2)/C_5^A(q^2)$ as a pure monopole term
\be
\frac{C_6^A(q^2)}{C_5^A(q^2)} = \frac{m_N^2}{m_{\pi}^2 -q^2}
\quad.
\label{monopole}
\ee
The aim here  is to calculate the dominant axial $C_5^A(q^2)$, $C_6^A(Q^2)$, as well as 
the pseudo-scalar $G_{\pi N \Delta}(Q^2)$ form factor
and examine the validity of the GT relations within the dynamical DWF framework, using both the coarse and fine DWF lattices. 

\subsection{\label{sec:domAXC5C6} The dominant axial $C^{A}_{5},$ $C^{A}_{6}$ transition form factors}

The extraction of the axial transition form factors requires data from two sets of the optimal $\Delta$ sinks, namely $S_1$ and $S_2$,
which are introduced in section II, for the local isovector axial-vector current insertion $A^3_\mu(x)$.
The corresponding expressions for the large Euclidean time separation ratios are:
%\small
%
\begin{align}
 S_1^A(\mathbf{q};j) 
=& B
\Bigg[-\frac{C^A_3}{2}\bigg\lbrace (E_N-2m_\Delta+m_N)  +
\left(\sum_{k=1}^3 p^k\right)\frac{p^j}{E_N+m_N} \bigg\rbrace  \nn \\
&- \frac{m_\Delta}{m_N}(E_N-m_\Delta)C^A_4
+m_N C^A_5-\frac{C^A_6}{m_N}p^j\left(\sum_{k=1}^3 p^k\right)\Bigg] ,
\label{S1A}
\end{align}
for spatial components $j=1,2,3$ of the axial current, and
\begin{align}
S_1^A(\mathbf{q};4) 
= -iB
 \sum_{k=1}^3 p^k\Bigg[C^A_3+\frac{m_\Delta}{m_N}C^A_4
+\frac{E_N-m_\Delta}{m_N}C^A_6\Bigg],
\label{S1A_4}
\end{align}
for the temporal component. Since the four form factors are not completely decoupled  by the above relations, we also 
employ the optimal $\Delta$ sink $S^A_2$ given in the plateau by
\begin{align}
\vspace{-0.3cm}
S_2^A(\mathbf{q};j) = 
 i \frac{3 A}{2}\Bigg[\left(\sum_{k=1}^3 p^k\right) 
\left(\delta_{j,1}(p^2-p^3)+
\delta_{j,2}(p^3-p^1)+\delta_{j,3}(p^1-p^2)\right)C^A_3\Bigg] ~,
\label{S2A}
\end{align}
\normalsize
valid for spatial components $j=1,2,3$. The kinematical factors $A$ and $B$ are given by
\begin{align} \label{KinFactorsAX}
A =\frac{B}{(E_N+m_N)} ~~~~,~~~~ B =\sqrt{\frac{2}{3}}\frac{\sqrt{\left(E_N+m_N\right)/E_N}}{3m_N} ~.
\end{align}
Data from $S^A_1$ and $S^A_2$ determine all four form factors $C^A_3$, $C^A_4$, $C^A_5$ and  $C^A_6$ at each value of $Q^2$  
in a simultaneous overconstrained analysis. $Z_A$ is required to renormalize the axial vector operator. This has been
computed by the UKQCD-RBC and LHP collaborations for both ensembles  \cite{Aoki:2007xm,Allton:2008pn,Syritsyn:2009mx}. The values provided in 
Table~\ref{Table:params_DWF_hybrid} confirm that $Z_V = Z_A + O(a^2)$ in the chiral limit, as expected for the manifestly chiral DWF action.

\begin{figure}[htb]
\centering
\subfigure[][]{\label{fig:C5ff}%
\includegraphics[width=0.48\linewidth,height=0.48\linewidth]{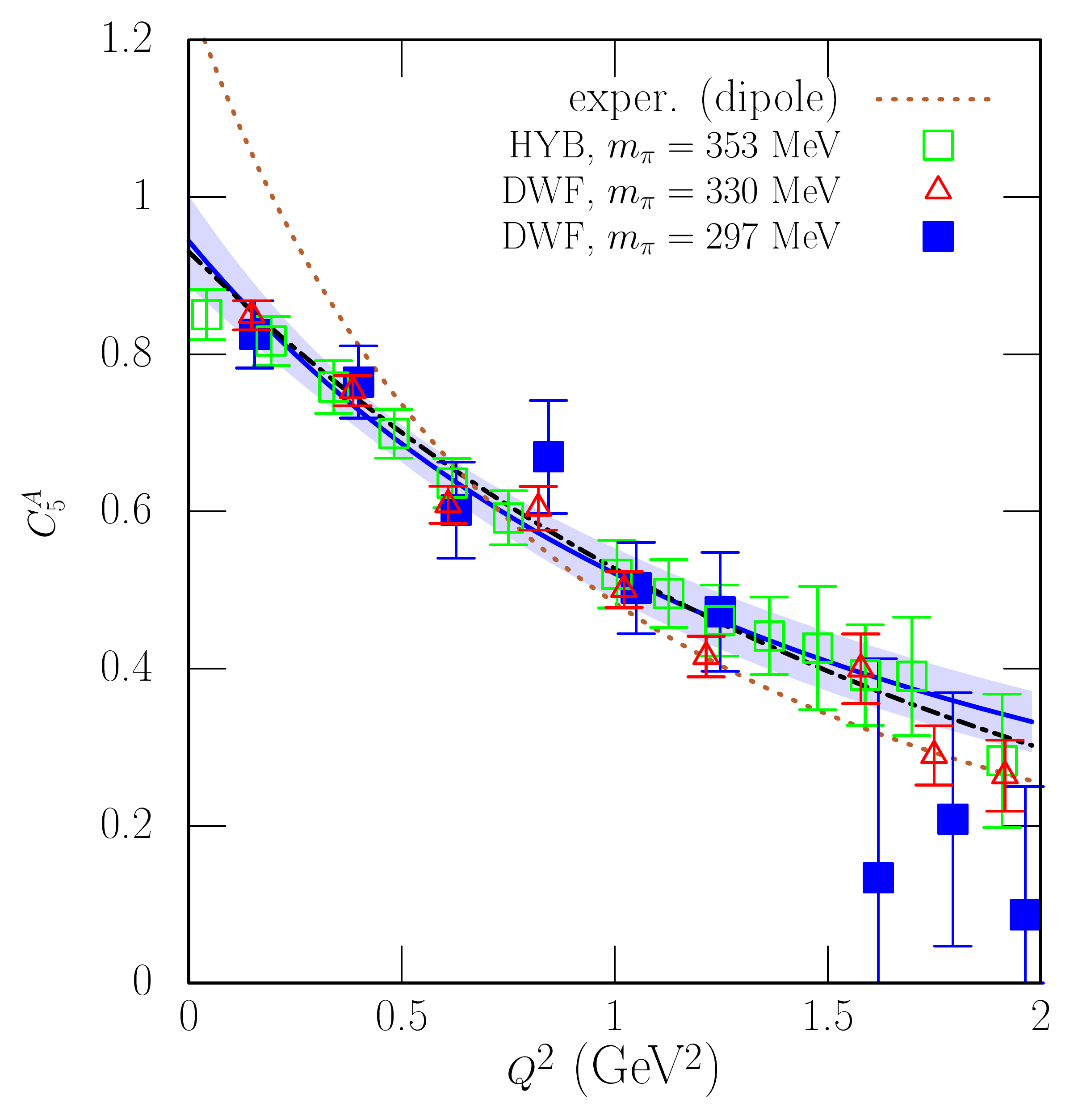} }
\hspace{-3pt}
\subfigure[][]{\label{fig:C6ovC5ff}%
\includegraphics[width=0.48\linewidth,height=0.48\linewidth]{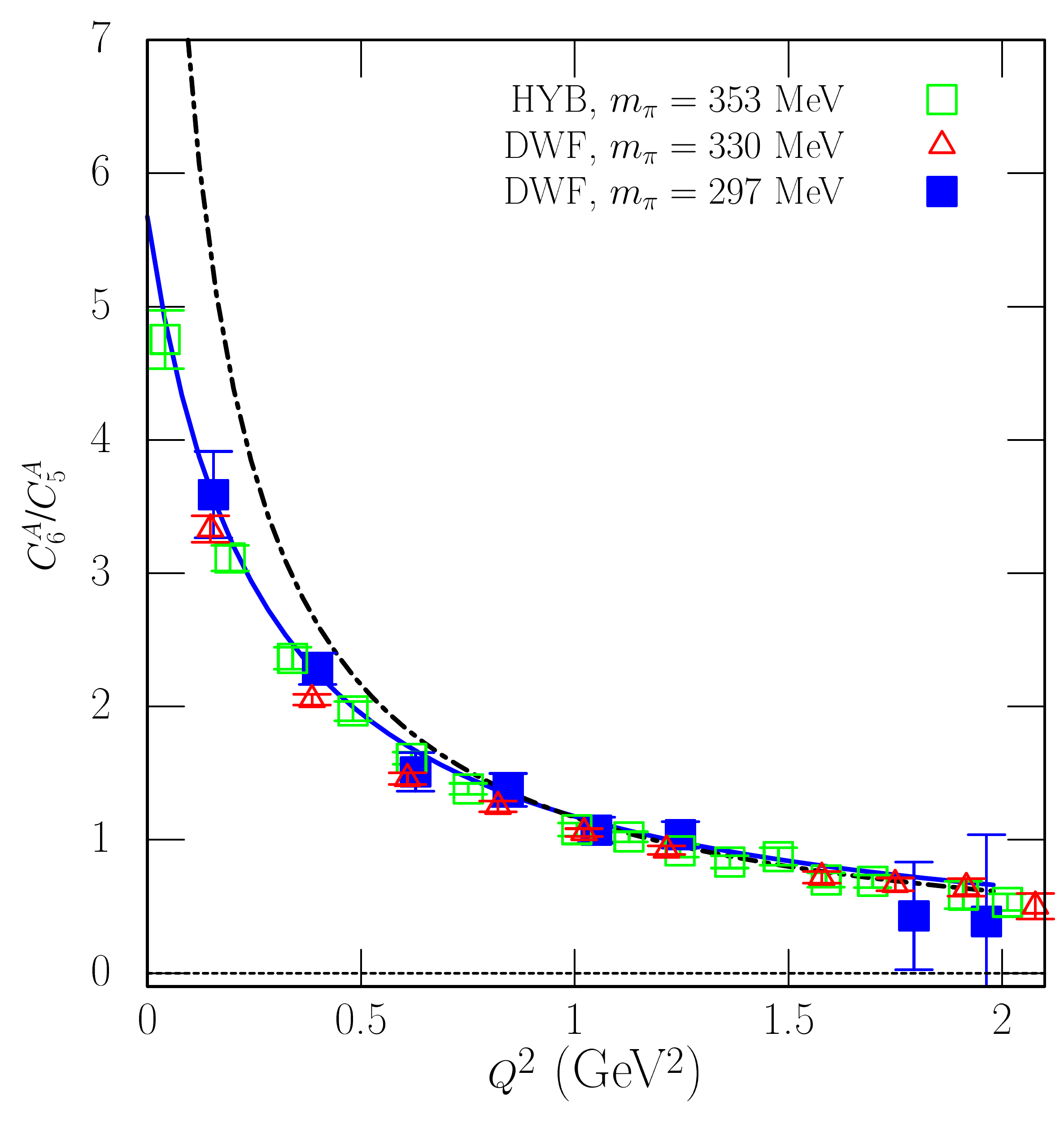} }
\caption{Plot \subref{fig:C5ff} shows the $Q^2$-dependence of the axial form factor $C^{A}_{5}$ extracted from the coarse and fine 
DWF lattices. The corresponding mixed action results~\cite{Alexandrou:2007xj} have also been included. The solid blue (dashed black) line is from 
the dipole (exponential) fit for to the fine DWF lattice results. Note that the error band corresponds to the dipole fit. The dotted brown line is the dipole fit to the experimental data.
The ratio $C_6^A/C_5^A$ versus $Q^2$ is plotted in \subref{fig:C6ovC5ff}. The dashed black line refers
to the fine DWF lattice results and is the pion pole dominance prediction of Eq.~(\ref{monopole}).
The solid blue  line is a fit to a monopole form $c_0/(1+Q^2/m^2)$.
}\label{fig:C5-C6}
\end{figure}

The results for the axial dominant form factor $C^{A}_{5}$ from the two DWF lattices  considered in this work  are 
presented in Fig.~\ref{fig:C5ff}
and are in good agreement with the results obtained from the mixed action approach at $m_{\pi}=353$~MeV~\cite{Alexandrou:2007xj}.  
The $Q^2$ dependence is well described by two-parameter dipole (solid line) and exponential (dashed line) forms
$d_0/(1+ Q^2/m_A^2)^2$, $\tilde{d}_0\exp(- Q^2/ \tilde{m}_A^2)$, respectively, which are almost indistinguishable in the plot.
The  fitted values for  $C^A_{5}(0)\equiv d_0$ (or $\tilde{d}_0$ of the exponential form) and the 
corresponding axial mass $m_{A}\, (\tilde{m}_A)$ are given in Table~\ref{Table:fitparams}. 
 In the same figure, we also show a dipole fit  to the available experimental data~\cite{Kitagaki:1990vs} which  determine  
an axial mass within the range of values of  $m_A \sim  0.85 - 1.1$~GeV~\cite{Kitagaki:1990vs,Hernandez:2010}, obtained from the pure dipole parameterization. As in the case of $G_{M1}(Q^2)$, we observe a flatter slope for 
the lattice data, reflected in the larger value of the axial mass $m_A$
extracted from the lattice results. 

%
%axial--fit params table
%
\begin{widetext}
\begin{center}
\begin{table}[h]
\scriptsize
\begin{tabular}{cccccccccc}
 \hline \hline  
$m_\pi$~[GeV] &  $m_A$ [GeV] & $d_0$ & $\tilde{m}_A$ [GeV]&  $\tilde{d}_0$ &$m$ [GeV] & $c_0$   &  $\Delta^\prime$ & \multicolumn{2}{c}{$g_{\pi N\Delta}$} \\ 
\hline
\multicolumn{8}{l}
{coarse $N_F=2+1$ DWF}  & ($K$)  & ($\alpha^{\prime}$)\\
0.329(1) & 1.588(70) & 0.970(30)   & 1.262(36)  & 0.940(21) & 0.509(15) &  5.132(204) & 0.030(5) & 9.525(168)&  13.936(588)\\
\hline
\multicolumn{8}{l}
{fine $N_F=2+1$ DWF}\\
0.297(5) & 1.699(170) &  0.944(58)  & 1.314(98)  & 0.927(46) & 0.507(33) & 5.756(516) & 0.037(6) & 8.444(491)& 16.257(867)\\
\hline
\multicolumn{10}{l}
{Hybrid action}\\
0.353(3) & 1.795(40) & 0.903(11)   & 1.386(18)  & 0.888(8) & 0.496(10) &  5.613(150) & 0.019(11) & 9.323(219)& 11.446(617)\\
\hline
\end{tabular}
\normalsize
\caption{The first column gives the pion mass in GeV. The
second  and third columns provide the dipole fit parameters $m_A$ and $d_0$ extracted from
fitting $C_5^A$ to $d_0/(1+Q^2/m^2_{A})^2,$ the fourth
and fifth  columns the corresponding parameters obtained from the use of an exponential
ansatz $\tilde{d}_0 \exp(-Q^2/\tilde{m}_A^2),$      the sixth and seventh
columns the fit parameters $m$ and $c_0$ extracted from  fitting the ratio
$C_6^A/C_5^A$ to a monopole form $c_0/(1+Q^2/m^2)$ for the  N--to--$\Delta$ process. 
The eighth and ninth columns show the calculated values of the fit parameters  $\alpha^{\prime}$ and  $\Delta^{\prime}$ 
defined in the linear fit of Eq.~(\ref{fit_str8_G}). The last two columns give
the predicted values of the strong coupling constant 
$g_{\pi N\Delta}\equiv G_{\pi N \Delta}(0).$ The first value
of the strong coupling constant is determined using
the fit function of Eq.~(\ref{fit_G}),
while the second uses the  linear fit based  
on Eq.~(\ref{fit_str8_G}), which is exactly equal to $\alpha^{\prime}.$ }
\label{Table:fitparams}
\end{table}
\end{center}
\end{widetext}
\normalsize

In Fig.~\ref{fig:C6ovC5ff} we show the
ratio $C_6^A/C_5^A, $ since ratios of this type are expected to be less prone to lattice 
artifacts. The dashed black line shows the pion pole dominance
prediction of Eq.~(\ref{monopole}), where for  $m_N$ and $m_\pi$ we use
the lattice extracted values that correspond to the fine DWF lattice. 
The predicted curve does not  describe 
the data at low-$Q^2$ i.e., in the regime where the strong pion cloud effects
are expected to be present. However, the fit to the monopole form
$ c_0/(1+ Q^2/m^2) $
describes satisfactorily the ratio yielding a heavier mass parameter $m$ than the lattice value of the  pion mass (see Table~\ref{Table:fitparams}). 
Such behavior has been observed also for the hybrid and quenched Wilson actions~\cite{Alexandrou:2007xj}.

\begin{figure}[htb]
\begin{center}
\hspace{-0.2cm}\includegraphics[width=0.65\linewidth,height=0.65\linewidth]{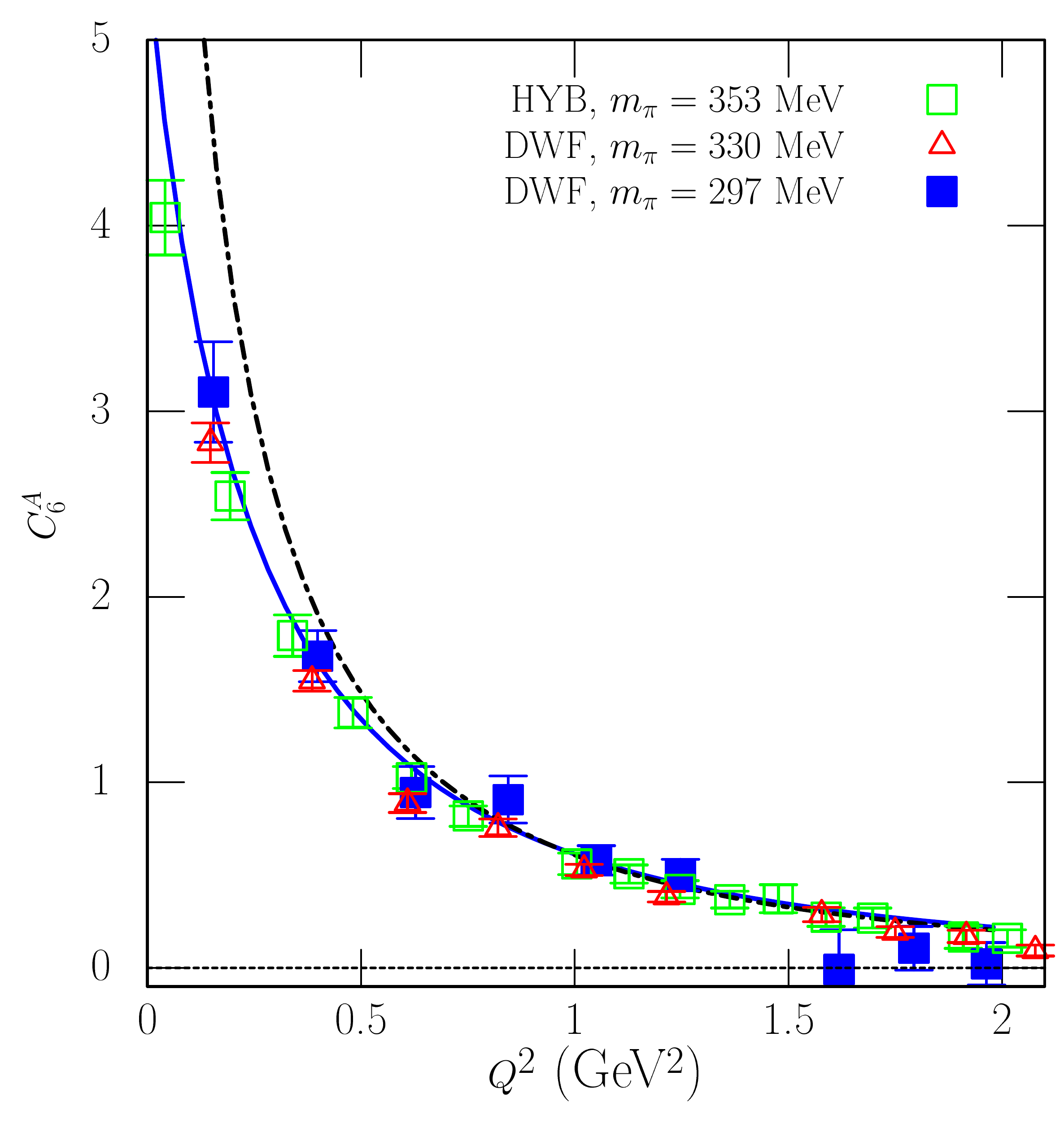} 
\vspace*{-0.5cm}
\caption{
Lattice results for $C^A_{6}$ are shown as a function of $Q^2$. The solid blue 
line is the fit to the form of Eq.~(\ref{C6_good_fit}), while the dashed black line corresponds to the form  $C^A_{5}\big(\frac{m^{2}_{N}}{m^{2}_{\pi}+Q^2}\big).$ Note that  for the latter fit, the $C^A_{5}$ factor is described by the dipole fit parameters.}
\label{fig:C6ff}
\end{center}
\end{figure}

The lattice results for the $C^A_{6}$ are plotted on Fig.~\ref{fig:C6ff}. 
The curve shown (solid line) in the figure corresponds to the form
\be
\frac{d_0~c_0}{(1+ Q^2/m_A^2)^2 (1+Q^2/m^2)},
\label{C6_good_fit}
\ee
where  $c_0$ and $m$ are the parameters of the  monopole term given in Eq.~(\ref{monopole}) that are expected to  describe well 
the $C^A_6/C^A_5$  ratio provided the pion pole dominance is applicable. The form described by the expression of Eq.~(\ref{C6_good_fit}), seems to provide the best fit to the fine DWF data. On the other hand, $C^A_{6}$ is related to the $C^A_{5}$ form factor through the expression 
\[C^{A}_{6}(Q^2)=C^A_{5}(Q^2)\ \frac{m^{2}_{N}}{m^{2}_{\pi}+Q^2}.\]
The curve that corresponds to the dashed line is obtained from fitting the fine DWF data to this form. In this case  
$C^{A}_{5}$ is being described by the dipole form shown in Fig.~\ref{fig:C5ff}, while the nucleon and pion masses are the lattice evaluated ones.

\subsection{\label{sec:gpND} The Pseudo-scalar transition form factor and Goldberger-Treiman relation}

The pseudo-scalar form factor $G_{\pi N \Delta}(Q^2)$,  defined via the matrix element given in Eq.~(\ref{gpiND}), is 
extracted directly from the optimized linear combination $S_1$   
with the pseudo-scalar current 
operator insertion of Eq.~(\ref{currents}).
In the large Euclidean time limit where
only the nucleon and $\Delta$ states dominate  the corresponding ratio yields
\beq  
 S^P_1({\bf q}\; ;\; \gamma_5) = 
 \sqrt{\frac{2}{3}}\sqrt{\frac{E_N+m_N}{E_N}} 
   \left[\frac{q_1 + q_2 + q_3}{6 m_N} \frac{f_\pi m_\pi^2} 
{2 m_q (m_\pi^2 + Q^2)} \right]\; G_{\pi N \Delta} (Q^2)\quad.
\label{S1P}
\eeq 

Notice that the extraction of $G_{\pi N \Delta}$ from the above equation requires  knowledge of the quark mass
$m_q$ and the pion decay constant, $f_{\pi}$, on the given ensembles.
Calculation of $f_{\pi}$ requires the two-point functions of the axial-vector current $A_4^3$ with local-smeared (LS) 
and smeared-smeared (SS) quark sources,
\beq
C^{A}_{LS}(t) 
 = \sum_{{\bf x}}  \; \langle \Omega
|\;T\;\left( A_4^3 ({\bf x},t)
 \tilde{A}^3_4 ({\bf 0},0)\right) \; |  \Omega\;\rangle 
\eeq
(and similarly for $C^A_{SS}$), where $A_4^3 ({\bf x},t)$ denotes the local operator and $\tilde{A}^3_4 ({\bf x},t)$ the 
smeared operator. The pion-to-vacuum matrix element 
\be 
\langle 0|A_\mu^a(0)|\pi^b(p)\rangle = i f_\pi p_\mu \delta^{ab}
\label{pion decay}
\ee
is extracted from the two-point functions $C^A_{LS}$ and $C^A_{SS}$ and
\beq
f_\pi^{\rm eff}(t) = Z_A \sqrt{\frac{2}{m_\pi}}
\frac{C^{A}_{LS} (t)}{\sqrt{C^{A}_{SS}(t)}}
\; e^{m_\pi t/2} \quad.
\label{fpi}
\eeq
yields  $f_{\pi}$
in the large Euclidean time limit.

The renormalized quark mass $m_q$ is determined from the AWI, via two-point functions of the pseudoscalar density with either
local ($P^3$) or smeared ($\tilde{P}^3$) quark fields,
\beq
C^{P}_{LS}(t) 
 = \sum_{{\bf x}}  \; \langle \Omega
|\;T\;\left( P^3 ({\bf x},t)
 \tilde{P}^3 ({\bf 0},0)\right) \; |  \Omega\;\rangle ~,
\eeq
(and similarly for $C^P_{SS}$).
The effective quark mass is defined by
\beq
m_{\rm eff}^{\rm AWI}(t) =\frac{m_\pi}{2}\frac{Z_A}{Z_P}
\frac{C^{A}_{LS} (t)}{C^{P}_{LS}(t)}
\sqrt{\frac{C^{P}_{SS} (t)}{C^{A}_{SS}(t)}} ~.
\label{meff}
\eeq
and its plateau value yields $m_q$.
Note that $Z_P$ will be needed only if ones wants $m_q$ alone. Since $Z_P$ enters also Eq.~(\ref{S1P}) it  cancels --as does
 $Z_A$ since it comes with $f_{\pi}$--  and 
therefore $G_{\pi N \Delta}$ is extracted directly from ratios of lattice three- and
two-point functions without prior knowledge of either $Z_A$ or $Z_P$. 
We also note that the quark mass computed 
through~(\ref{meff}) includes the
effects of residual chiral symmetry breaking from the finite extent $L_5$ 
of the fifth dimension. These effects are of the order of $60\%$ for the 
coarse ensemble and $17 \%$ for the fine ensemble.
 Chiral symmetry breaking affects the PCAC relations and
 therefore the value of $G_{\pi N \Delta}$ 
through Eq.~(\ref{S1P}). 

\begin{figure}[htb]
\centering
\subfigure[][]{\label{fig:GPovC5ff}%
\includegraphics[width=0.48\linewidth,height=0.48\linewidth]{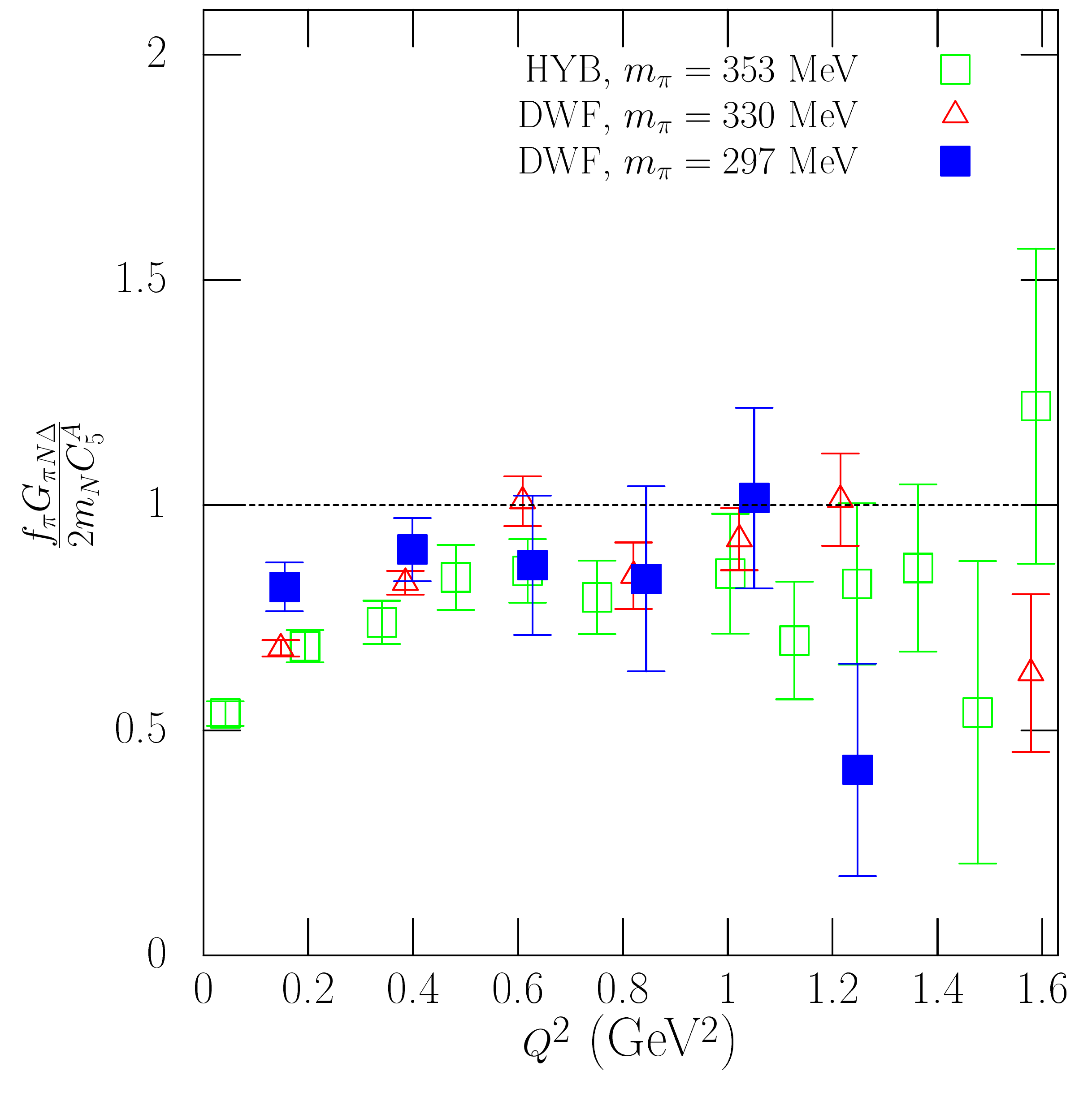} }
\hspace{-3pt}
\subfigure[][]{\label{fig:GPovC6ff}%
\includegraphics[width=0.48\linewidth,height=0.48\linewidth]{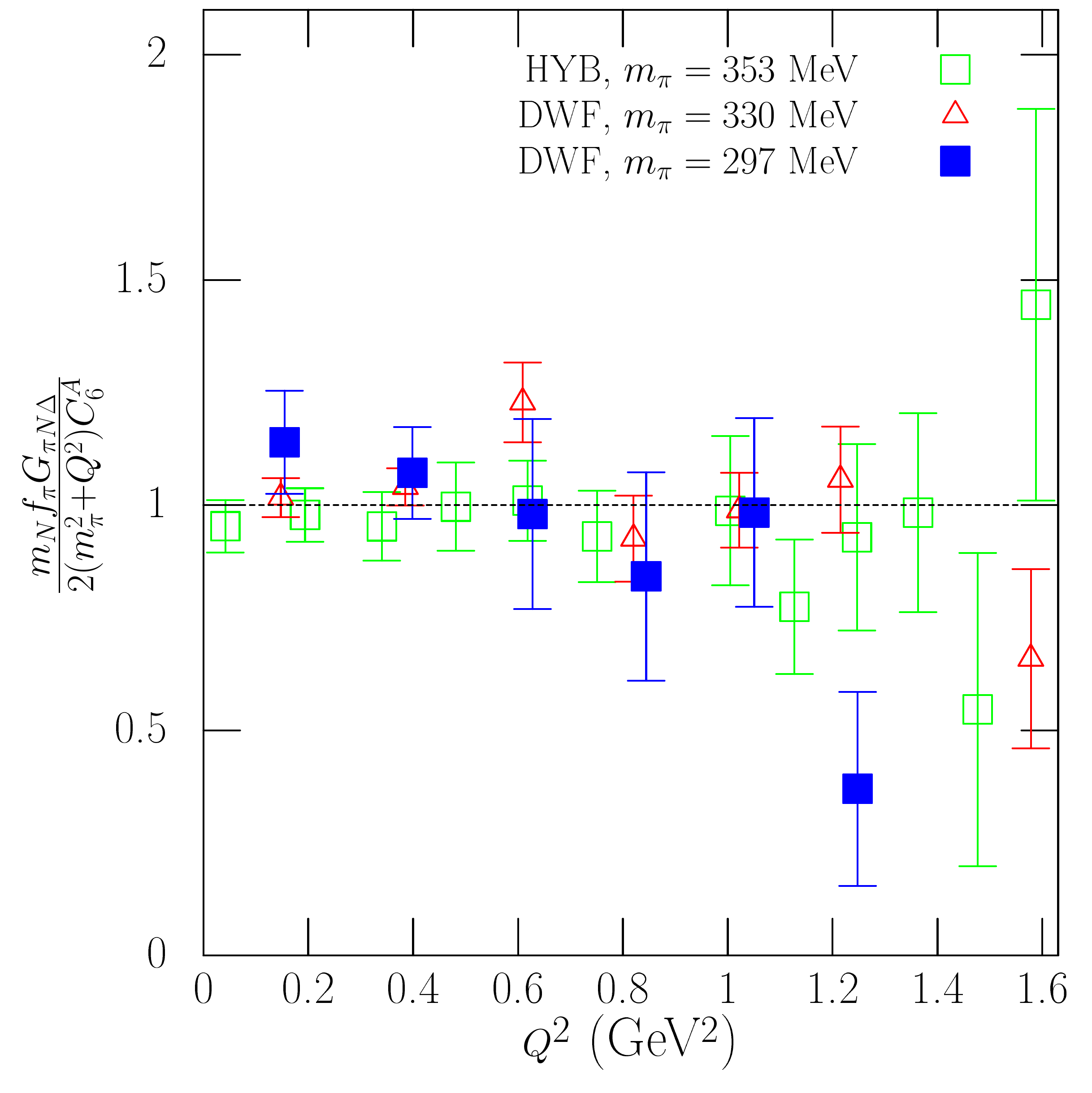}}
\caption{In \subref{fig:GPovC5ff} we plot the ratio of  Eq.~(\ref{ratioGPovC5}) as a function of $Q^2$ as a validity test of the GT relation.
Similarly, in plot \subref{fig:GPovC6ff}   ratio  of  Eq.~(\ref{ratioGPovC6}) that relates to the validity of Eq.~(\ref{GP}).}\label{fig:GPovC5C6}
\end{figure}

The ratio 
\begin{align}\label{ratioGPovC5}
\frac{f_\pi G_{\pi N\Delta}(Q^2)}{2m_N C_5^A(Q^2)}
\end{align}
is depicted in Fig.~\ref{fig:GPovC5ff}. 
 It should 
be \emph{unity} if  the off-diagonal Goldberger-Treiman relation of
Eq.~(\ref{GTR}) is satisfied, which in turn requires that PCAC holds 
exactly at the pion masses simulated in these ensembles.
Deviations from this relation are 
seen in the low-$Q^2$ regime.
For the fine ensemble considered in this study, the 
deviations from unity are less severe.
At momentum transfers, of about $Q^2 \gtrsim 0.5$~GeV$^2$,  the relation is at least approximately satisfied and it is consistent among all actions considered here.

 Pion pole dominance relates $C_6^A$ to $C_5^A$
through Eq.~(\ref{monopole}), which is very well satisfied by the lattice
data for all the  actions employed in this work (see also Fig.~\ref{fig:C6ovC5ff}). This agreement is also evident from Fig.~\ref{fig:GPovC6ff}, where the ratio
\begin{align}\label{ratioGPovC6} 
\frac{m_N f_\pi G_{\pi N\Delta}(Q^2)}{ 2(m_{\pi}^2 + Q^2)C_6^A(Q^2)}
\end{align} 
is consistent with unity.

\begin{figure}[htb]
\begin{center}
\hspace{-0.2cm}\includegraphics[width=0.65\linewidth,height=0.65\linewidth]{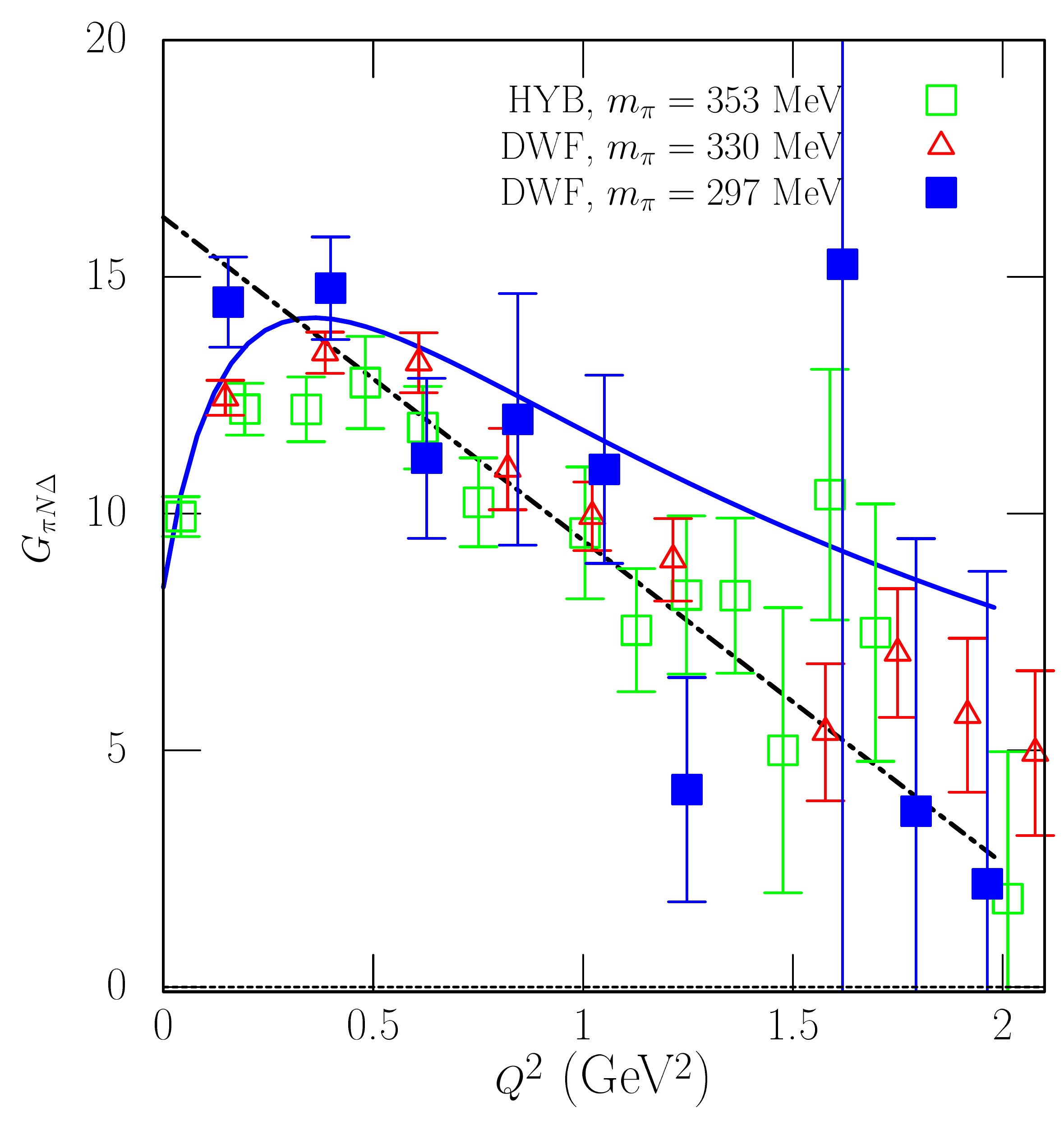} 
\vspace*{-0.5cm}
\caption{
The plot shows the $Q^2$-dependence of the pseudo-scalar transition form factor
$G_{\pi N \Delta}$. The solid blue line is a fit to pion pole dominance form of 
Eq.~(\ref{fit_G}) for the fine DWF ensemble. The dashed black line is the linear fit given by Eq.~(\ref{fit_str8_G}). 
The strong coupling constant $g_{\pi N \Delta}$ is the value of $G_{\pi N \Delta}$ at $Q^2 = 0$.}\label{fig:GPND}
\end{center}
\end{figure}

In Fig.~\ref{fig:GPND} we compare results on
 $G_{\pi N \Delta}(q^2)$ using the dynamical DWF lattices to the results
obtained from the hybrid scheme taken from Ref.~\cite{Alexandrou:2007xj}.
There is an agreement for  $Q^2>0.5$~GeV$^2$ whereas for lower $Q^2$ values
the fine DWF data appear to be higher than the data from the other two 
lattices. 
The solid line is a one-parameter fit form  to the fine DWF data  
\begin{align}
\label{fit_G}
G_{\pi N \Delta}(Q^2) &=K \>\frac{(Q^2/m_\pi^2+1)}{(Q^2/m_A^2+1)^2(Q^2/m^2+1)}, 
\end{align}
which is expected assuming the validity of Eq.~(\ref{monopole}). 
The fit parameter $K$ provides an estimate of the strong coupling
$g_{\pi N \Delta}$ at   $Q^2=0$. In addition, we fit to 
the ansatz
\begin{align}\label{fit_str8_G}
G_{\pi N\Delta}(Q^2) = \alpha^{\prime}  \biggl(1 -\Delta^{\prime} \frac{Q^2}{m_\pi^2}\biggr),
\end{align}
shown by the dashed line. The fit parameters  are provided in Table~\ref{Table:fitparams}. 
As can be seen, despite the fact that  both fits describe sufficiently well the data for $0.5 \lesssim Q^{2} \lesssim 1.5$~GeV$^2$, they yield quite  different values at $Q^2=0$
prohibiting a reliable evaluation of
$g_{\pi N \Delta}$. 
 Clearly, in order to achieve this goal, a better  
understanding of the behavior at low-$Q^2$ is required, since this quantity is
sensitive to pion loop effects that maybe affected by  lattice artifacts
such as the finite-$L_5$ extent.

\begin{figure}[htb]
\begin{center}
\hspace{-0.2cm}\includegraphics[width=0.65\linewidth,height=0.65\linewidth]{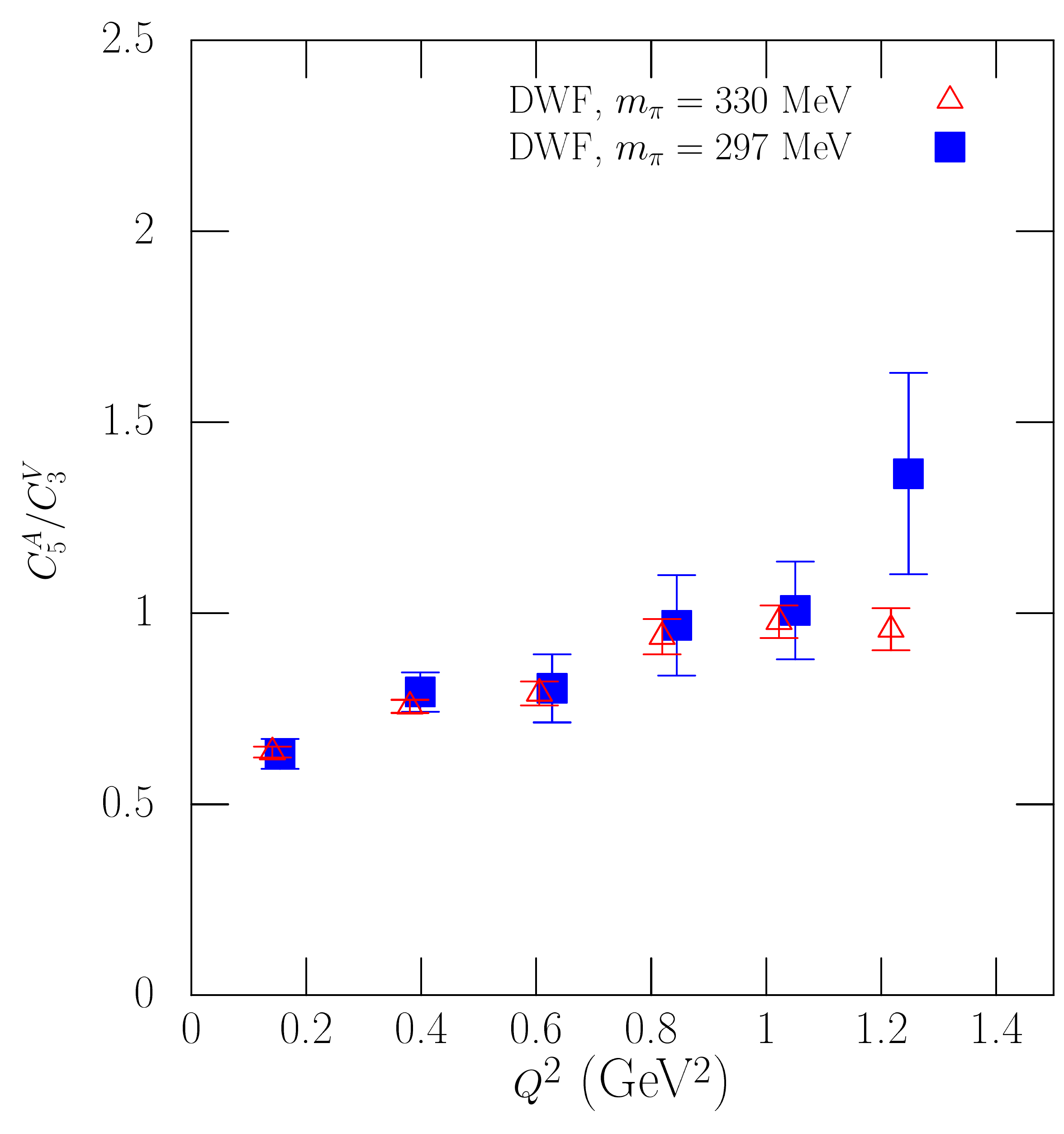} 
\vspace*{-0.5cm}
\caption{
The plot describes the $Q^2$-dependence of the ratio $C^{A}_{5}/C^{V}_{3}$. The results shown are those extracted from both DWF lattices considered in this work.}\label{fig:C5ovC3v}
\end{center}
\end{figure}

Finally, from our lattice results we can predict the currently unmeasured ratio $C^{A}_{5}/C^{V}_{3}$, which is an important first approximation to the parity violating asymmetry.  Its dependence in $Q^2$ is depicted in Fig.~\ref{fig:C5ovC3v}. From the plot we can see a very good agreement between the coarse and fine DWF data, at least in the range up to $Q^2 \sim 1.0$~GeV, indicating that there are no lattice cut-off effects regarding this quantity. It is also evident from the plot that at $Q^2=0$ the ratio is expected to have a non-zero value. It is noted that  $C^{V}_{3}$ is computed from the relationship
\begin{align}
C^{V}_{3}=\frac{3}{2}\ \frac{m_{\Delta}(m_{N}+m_{\Delta})}{2(m_{N}+m_{\Delta})^{2}+Q^2}\ (G_{M1}-G_{E2}) 
\label{eq:C3v}
\end{align}
 and is therefore dominated by $G_{M1}$. As both $C_5^A$ and 
$G_{M1}$ lack
chiral effects near the origin, the ratio $C^{A}_{5}/C^{V}_{3}$ is expected 
to be less sensitive to such effects.
The present results for $C^{A}_{5}/C^{V}_{3}$ are also consistent within statistics with the results reported earlier in Ref.~\cite{Alexandrou:2006mc}.

\section{\label{sec:conclusions} Conclusions}

The nucleon to $\Delta$ electromagnetic, axial and pseudoscalar transition 
form factors are calculated using $N_f=2+1$ dynamical domain wall fermions for pion masses
of 330 MeV and 297 MeV for  $Q^2$ values up to about  2~GeV$^2$. 
There is qualitative agreement between results obtained in the
unitary theory and corresponding results obtained using valence
domain wall quarks on a staggered sea.
The momentum dependence of the dominant magnetic dipole, $G_{M1}$, and axial, $C_5^A,$ form factors are well described by dipole forms.
They both show a 
slower fall-off with $Q^2$ than the comparison to the experimental data, a fact that is reflected in the heavier dipole masses that 
fit the lattice data. Pion cloud effects are expected
to dominate the low-$Q^2$ dependence and therefore simulations
with pion mass below 300 MeV are required in order to allow the evaluation of
such effects from first principles.

The phenomenologically interesting sub-dominant electromagnetic quadrupole form factors $G_{E2}$ and $G_{C2}$ have been  
calculated in the case of the fine DWF lattice using  the coherent sink technique  in order to increase  the statistical accuracy.
The results confirm a non-zero value at low $Q^2 \le 1~{\rm GeV}^2$.
The
EMR and CMR ratios are almost $Q^2$ independent. The EMR 
values are in agreement with the experiment, whereas 
the strength of the CMR  is underestimated.
This can be understood in chiral effective theory, which
predicts 
different chiral behavior for   the two quantities.
The non-zero values calculated in  QCD  are in accord with the experimental 
determinations~\cite{Mertz:2001,Joo:2001,Sparveris:2004jn,Sparveris:2006uk,Stave:2006ea,Papanicolas:2003zz} and confirm  a deviation from spherical symmetry in the Nucleon-$\Delta$ system. 

The axial transition form factor $C^{A}_{6}$  is dominated by chiral symmetry breaking dynamics, which is directly reflected in the
pion pole dominance.
 In addition, the pseudoscalar form
factor $G_{\pi N \Delta}$ is computed and the non-diagonal Goldberger-Treiman relation, which is a direct consequence of  
PCAC is shown to be well satisfied by the lattice data, especially 
for the lowest mass on the fine DWF ensemble. Pure monopole dependence of the 
$C^A_6/C^A_5$ ratio is well satisfied, but with monopole masses considerably heavier than the corresponding lattice pion masses.
The low-$Q^2$ dependence of $G_{\pi N \Delta}$ appears to be non-trivial and the extraction of the phenomenological strong
$\pi-N-\Delta$ coupling, $g_{\pi N \Delta}$, requires careful understanding of the matrix element systematics, since it will be sensitive to
both chiral and lattice cutoff effects. 

In conclusion, the $N-\Delta$ transition yields valuable information that is complementary to nucleon and Delta form factors.  Also, since the transition is isovector, it provides an opportunity to assess the importance of disconnected quark loop effects. Furthermore, it provides constraints on the low energy constants that enter the chiral effective description of hadron properties.
This work, utilizing dynamical chiral fermions corresponding to pion masses of 297~MeV and 330~MeV, together with related calculations of nucleon and Delta form factors, is a significant advance in the quest to understand from first principles how the  closely related structure of the nucleon and Delta arise from QCD.  The outstanding challenge for the future is to extend these 
calculations to the physical pion mass and reduce statistical and systematic errors to the level of a few percent. It is an  appealing challenge for Lattice QCD to perform precise calculations for pion masses that approach the physical point with all systematics under control. Simulations with pions almost at its physical value will soon become available
and it will be important to continue the investigation of 
these quantities. 
\newline

\centerline{\bf Acknowledgments}

This research was partly supported by the Cyprus Research Promotion Foundation (R.P.F) under contracts No. $\mathrm{\Pi}$ENEK/ENI$\mathrm{\Sigma}$X/0505-39 and No. EPYAN/0506/08 and by the U.S. Department of Energy under Grant No. DE-FG02-94ER-40818. The authors would also like to acknowledge the use of dynamical domain wall fermions configurations provided by the RBC-UKQCD collaborations, the forward propagators provided by the LHPC and the use of Chroma software~\cite{Edwards:2004sx}.
\bibliography{N2Delta_ref}
\newpage
%

%-------------------------------------------------------------------
% Tables for EM and AXIAL NDelta data from DWF (a^-1=1.73,2.34 GeV) 
%-------------------------------------------------------------------
\section{Appendix}
%coarseDWF-GM1
\begin{table}[h]
\begin{tabular}{cc}
\hline 
$Q^2$ (GeV$^2$) & $G_{M1}$ \\ 
\hline
\multicolumn{2}{c}{DWF ($N_f=2+1$), $a^{-1}=1.73$~GeV, $m_\pi=330$~MeV} \\
\hline  
\hline
0.141 & 1.581(40)  \\
0.380 & 1.198(32)  \\
0.605 & 0.933(33)  \\
0.819 & 0.786(39)  \\
1.022 & 0.641(30)  \\
1.217 & 0.545(33)  \\
1.584 & 0.449(50)  \\
1.757 & 0.369(42)  \\
1.925 & 0.332(51)  \\
2.088 & 0.238(48)  \\
2.247 & 0.204(99)  \\
\hline
\hline
\end{tabular}
\caption{Coarse DWF results for $G_{M1}$, their  $Q^2$-dependence and the corresponding (form factor) jackknife statistical errors.  
}
\label{Table:coarseDWFGM1data}
\end{table}
%fineDWF
\begin{table}[h]
\begin{tabular}{cccccc}
\hline 
$Q^2$ (GeV$^2$) & $G_{M1}$ & $G_{E2}$  & EMR $(\%)$ & $G_{C2}$ & CMR $(\%)$ \\ 
\hline
\multicolumn{6}{c}{DWF ($N_f=2+1$), $a^{-1}=2.34$~GeV, $m_\pi=297$~MeV} \\
\hline  
\hline
0.154 & 1.602(93) & 0.0508(344) & -3.118(2.064) & 0.249(142) & -2.748(1.595) \\
0.398 & 1.168(75) & 0.0146(208) & -1.129(1.686) & 0.122(98) & -2.624(2.144) \\
0.627 & 0.928(84) & 0.0156(259) & -1.528(2.749) & 0.006(124) & -3.145(4.036) \\
0.844 & 0.875(101) & 0.0441(375) & -5.246(4.259) & 0.158(105) & -6.439(4.348) \\
1.051 & 0.593(72) &0.0261(225) &4.263(3.707) & 0.186(67) & -12.490(4.742) \\
1.248 & 0.417(86) &0.0206(251) &4.874(5.781) & 0.197(74) & -20.847(8.769) \\
1.620 & 0.439(44) & & & & \\
1.802 & 0.224(159) & & & & \\
1.964 & 0.165(181) & & & & \\
\hline
\hline
\end{tabular}
\caption{DWF results for $G_{M1}$, $G_{E2}$, EMR $(\%)$, $G_{C2}$ and CMR $(\%)$ along with  their  $Q^2$-dependence shown in the first column. The errors shown are statistical jackknife errors.}
\label{Table:DWFEMdata}
\end{table}
%
%AXIAL
\begin{table}[h]
\begin{tabular}{cccc}
\hline 
$Q^2$~(GeV$^2$) & $C_5^A$ & $C_6^A$ & $G_{\pi N\Delta}$ \\
\hline 
\multicolumn{4}{c} {DWF ($N_f=2+1$), $a^{-1}=1.73$~GeV, $m_\pi=330$~MeV} \\
\hline  
\hline
0.141 & 0.849(19) & 2.831(106) & 12.446(371) \\
0.380 & 0.754(19) & 1.547(55) &  13.379(436) \\
0.605 & 0.608(24) & 0.888(51) & 13.187(633) \\
0.819 & 0.604(27) & 0.755(47) & 10.941(86) \\
1.022 & 0.500(23) & 0.528(29) & 9.943(727) \\
1.217 & 0.415(26) & 0.383(28) & 9.026(870) \\
1.584 & 0.399(44) & 0.287(38) & 5.379(1.448) \\
1.757 & 0.289(38) & 0.193(28) & 5.741(1.625) \\
1.925 & 0.263(45) & 0.169(33) & 7.055(1.359) \\
2.247 & 0.186(46) & 0.093(29) & 4.942(1.740) \\
\hline 
\multicolumn{4}{c} {DWF ($N_f=2+1$), $a^{-1}=2.34$~GeV, $m_\pi=297$~MeV} \\
\hline  
\hline
0.154 & 0.825(42) & 3.103(270) & 15.292(1.005) \\
0.398 & 0.764(46) & 1.680(138) & 15.601(1.145) \\
0.627 & 0.601(61) & 0.945(139) & 11.803(1.784) \\
0.844 & 0.669(72) & 0.907(127) & 12.672(2.807) \\
1.051 & 0.502(158) & 0.579(80) & 11.556(2.098) \\
1.248 & 0.472(76) & 0.501(85) & 4.040(2.504) \\
1.620 & 0.134(278) & -0.008(213) & 16.132(17.140) \\
1.802 & 0.208(161) & 0.105(117) & 3.924(6.088) \\
1.964 & 0.087(163) & 0.022(114) & 2.302(6.978) \\
2.128 & 0.097(384) & 0.084(275) &  -1.382(12.666)\\
%2.288 & 0.001(799) &  &  \\
\hline
\end{tabular}
\label{Table:DWFAXIALdata}
\caption{DWF results for $C^{A}_{5}$, $C^{A}_{6}$ and $G_{\pi N\Delta}$ along with  their  $Q^2$-dependence shown in the first column. 
The errors quoted are jackknife statistical errors.}
\end{table}
%

%%%%%%%%%%%%%%%%%%%%%%%%%%%%%%%%%%%%%%%%%%%%%%%%%%%%%%%%%%%%%%%%%%
\end{document}